%% file: rest-comparisons.tex
\pdfoutput=1 

\documentclass[11pt]{article}

\usepackage[square,numbers]{natbib}

\usepackage{geometry}
\usepackage{booktabs}
\usepackage{hyperref}
\usepackage{caption}
\usepackage{subcaption}

\usepackage{multirow}

\usepackage{xspace}
\usepackage{setspace}
\usepackage{enumerate}

\usepackage{censor}

\usepackage[T1]{fontenc} % needed for scaling fonts
\usepackage{lmodern}     % same as above, for scalable fonts

\usepackage{listings}

\usepackage{pdflscape}

\usepackage{microtype}
\DisableLigatures{}

\usepackage{xcolor}
\definecolor{codegreen}{rgb}{0.25,0.5,0.35}
\definecolor{codegray}{rgb}{0.5,0.5,0.5}
\definecolor{codepurple}{rgb}{0.6,0,0}
\definecolor{backcolour}{rgb}{0.95,0.95,0.92}
\definecolor{colorstring}{rgb}{0.5,0,0.35}
\definecolor{rltred}{rgb}{0.5,0,0}
\definecolor{rltgreen}{rgb}{0,0.5,0}
\definecolor{rltblue}{rgb}{0,0,0.5}
\definecolor{DarkGreen}{rgb}{0.00,0.60,0.00}
\definecolor{ScarletRed}{rgb}{0.80,0.00,0.00}
\definecolor{blizzardblue}{rgb}{0.67, 0.9, 0.93}
\definecolor{green-yellow}{rgb}{0.68, 1.0, 0.18}
\definecolor{dkgreen}{rgb}{0,0.6,0}
\definecolor{gray}{rgb}{0.5,0.5,0.5}
\definecolor{mauve}{rgb}{0.58,0,0.82}
\definecolor{lightgrey}{rgb}{0.90,0.90,0.90}
\definecolor{grey}{gray}{0.75}
\definecolor{light-gray}{gray}{0.80}

\lstdefinestyle{mystyle}{
	backgroundcolor=\color{backcolour},   
	commentstyle=\color{codegreen},
	keywordstyle=\color{colorstring}\bfseries,
	numberstyle=\ttfamily\color{codegray},
	stringstyle=\color{codepurple},
            basicstyle={\scriptsize\ttfamily},
	breakatwhitespace=false,         
	breaklines=true,                 
	captionpos=b,                    
	keepspaces=true,                 
	numbers=left,                    
	numbersep=2pt,                  
	showspaces=false,                
	showstringspaces=false,
	showtabs=false,                  
	tabsize=2
}
\newcommand{\bBOXRT}{{\sc bBOXRT}\xspace}
\newcommand{\evo}{{\sc EvoMaster}\xspace}
\newcommand{\RESTest}{{\sc RESTest}\xspace}
\newcommand{\RestTest}{{\sc RESTest}\xspace}
\newcommand{\RestCT}{{\sc RestCT}\xspace}
\newcommand{\RESTler}{{\sc RESTler}\xspace}
\newcommand{\Restler }{{\sc RESTler}\xspace}
\newcommand{\RestTestGen}{{\sc RestTestGen}\xspace}
\newcommand{\RestTestGenVT}{{\sc RestTestGen}\xspace}
\newcommand{\Schemathesis}{{\sc Schemathesis}\xspace}

\newcommand{\cwa}{{\emph{cwa-verification}}\xspace}
\newcommand{\featuresservice}{{\emph{features-service}}\xspace}
\newcommand{\gestao}{{\emph{gestaohospital-rest}}\xspace}
\newcommand{\indCS}{{\emph{ind0}}\xspace}
\newcommand{\languagetool}{{\emph{languagetool}}\xspace}
\newcommand{\ocvnrest}{{\emph{ocvn-rest}}\xspace}
\newcommand{\proxyprint}{{\emph{proxyprint}}\xspace}

\newcommand{\restncs}{{\emph{rest-ncs}}\xspace}
\newcommand{\restnews}{{\emph{rest-news}}\xspace}
\newcommand{\restscs}{{\emph{rest-scs}}\xspace}
\newcommand{\restcountries}{{\emph{restcountries}}\xspace}
\newcommand{\scoutapi}{{\emph{scout-api}}\xspace}

\newcommand{\cyclotron}{{\emph{cyclotron}}\xspace}
\newcommand{\diseasesh}{{\emph{disease-sh-api}}\xspace}
\newcommand{\jsrestncs}{{\emph{js-rest-ncs}}\xspace}
\newcommand{\jsrestscs}{{\emph{js-rest-scs}}\xspace}
\newcommand{\realworldapp}{{\emph{realworld-app}}\xspace}
\newcommand{\spacexapi}{{\emph{spacex-api}}\xspace}

\usepackage{pgfplots} 

\definecolor{ForestGreen}{RGB}{34,139,34}
\definecolor{asparagus}{rgb}{0.53, 0.66, 0.42}

\usepackage{boxedminipage}
\newenvironment{result}%
{\smallskip
	\noindent
	\let\emph=\textbf
	\begin{boxedminipage}{\columnwidth}\begin{center}\em}%
		{\end{center}\end{boxedminipage}%
}

\newboolean{showcomments}
\setboolean{showcomments}{true} % comment this line to deactivate comments

\ifthenelse{\boolean{showcomments}}{
	\newcommand{\nbc}[3]{
		{\colorbox{#3}{\bfseries\sffamily\scriptsize\textcolor{white}{#1}}}
		{\textcolor{#3}{\sf\small$\langle$\textit{#2}$\rangle$}}}
	
}{
	\newcommand{\nbc}[3]{}

}

\title{Open Problems in Fuzzing RESTful APIs:\\ A Comparison of Tools}
\author{Man Zhang, Andrea Arcuri\\Kristiania University College, Norway}
\date{}

\begin{document}

\maketitle

\begin{abstract}
RESTful APIs are a type of web services that are widely used in industry.
In the last few years, a lot of effort in the research community has been spent in designing novel techniques to automatically fuzz those APIs to find faults in them.
Many real faults were automatically found in a large variety of RESTful APIs.
However, usually the analyzed fuzzers treat the APIs as black-box, and no analysis of what is actually covered in these systems is done.
Therefore, although these fuzzers are clearly useful for practitioners, we do not know what are their current limitations and actual effectiveness.
Solving this is a necessary step to be able to design better, more efficient and effective techniques.
To address this issue, in this paper we compare seven state-of-the-art fuzzers on 18 open-source and one industrial RESTful APIs. 
We then analyzed the source code of which parts of these APIs the fuzzers fail to generate tests for. 
This analysis points to clear limitations of these current fuzzers, listing concrete challenges for the research community to follow up on.
\end{abstract}


%%%%%%%%%%%%%%%%%%%%%%%%%%%%%%%%%%%%%%%%%%%%%%%%%%%%%%%%%%%%%%%%%%%%%%%%%%%%
\section{Introduction}

RESTful APIs are widely used in industry.
Arguably, REST is the most common kind of web service, used to provide functionality (i.e., an API) over a network.
Many companies worldwide provide services via a RESTful API, like
Google~\cite{GoogleDriveAPI},
Amazon~\cite{AmazonGatewayAPI},
LinkedIn~\cite{LinkedInAPI},
Twitter~\cite{TwitterAPI},
Reddit~\cite{RedditAPI},
etc.
The website \emph{ProgrammableWeb}~\cite{ProgrammableWeb} currently lists more than 24 thousand APIs available on internet, where the vast majority of them is implemented with REST.
Furthermore, REST APIs are also commonly used when developing backend enterprise applications using a microservice architecture~\cite{newman2015building}.

Due to their wide use in industry, in recent years there has been a large interest in the research community to design novel techniques to automatically test this kind of applications (e.g.,~\cite{laranjeiro2021black,arcuri2017restful,restest2020,wu2022icse,restlerICSE2019,viglianisi2020resttestgen}).
Compared to other kinds of applications (e.g., data parsers) and type of testing (e.g., unit testing),
system testing of RESTful APIs has some unique challenges, like dealing with network communications (e.g., HTTP over TCP) and with accesses to databases (e.g., Postgres and MySQL).

In the literature, different techniques have been proposed, which have been able to automatically find many faults in existing APIs (e.g.,~\cite{laranjeiro2021black,arcuri2017restful,restest2020,wu2022icse,restlerICSE2019,viglianisi2020resttestgen}).
However, so far, most of the investigated techniques are black-box, applied on remote web services on the internet.
Although those fuzzers have been useful to detect several faults, it is unclear how effective they truly are at testing these APIs.
For example, most studies do not report any form of code coverage, as the system under tests (SUTs) are typically remote black-boxes with no access to their source code. 
For example, it can be possible that most found faults are simply in the first layer of input validation of these SUTs, with little-to-no code of their businesses logic executed~\cite{marculescu2022faults}. 
Without an in-depth analysis of which parts of the SUT code is not exercised by these fuzzers, it is not possible to understand their limitations.
This hinders further research development for novel techniques to achieve better results. 

To address these issues, in this paper we compare seven state-of-the-art fuzzers, namely (in alphabetic order)
\bBOXRT~\cite{laranjeiro2021black}, \evo~\cite{arcuri2017restful}, \RESTest~\cite{restest2020}, \RestCT~\cite{wu2022icse}, \RESTler~\cite{restlerICSE2019}, \RestTestGen~\cite{viglianisi2020resttestgen} and
\Schemathesis~\cite{hatfield2022deriving}.
We applied them on 13 RESTful APIs running on the JVM (i.e., 12 open-source and one industrial) and six open-source RESTful APIs running on NodeJS, for which we collected line coverage metrics using JaCoCo~\cite{JaCoCo} (for JVM) and c8~\cite{C8} (for NodeJS).
Experiments are carried both for black-box and white-box testing. 

This large set of experiments (which takes roughly 
71.2
days if run in sequence) enables us to analyze in details what are the current limitations of the state-of-the-art in fuzzing RESTful APIs.
These tools achieve different degrees of code coverage, but there are still many issues that need to be addressed to achieve better results.
These include for example how to deal with underspecified schemas, and how to deal with interactions with external services.

In this paper, we provide the following novel research contributions:
\begin{itemize}
\item  	We carried out one of the largest and most variegated to date empirical comparison of fuzzers for RESTful APIs.
\item Instead of using black-box metrics, we report actual coverage results on the employed APIs. 
\item    We provide in-depth analyses of the current challenges in this testing domain. 
\end{itemize}

The paper is organized as follows.
Section~\ref{sec:background} provides background info on the seven compared tools.
Related work is discussed in Section~\ref{sec:related}.
Section~\ref{sec:empirical} presents the details of the tool comparisons.
Section~\ref{sec:problems} provides the main scientific contribution of this paper, with an in-depth analysis of the current open-problems identified by our empirical study.
Threats to validity are discussed in Section~\ref{sec:threats}.
Finally, Section~\ref{sec:conclusions} concludes the paper.

%%%%%%%%%%%%%%%%%%%%%%%%%%%%%%%%%%%%%%%%%%%%%%%%%%%%%%%%%%%%%%%%%%%%%%%%%%%%
\section{Background: Used Tools}
\label{sec:background}

In this paper, we compare seven different fuzzers for RESTful APIs, namely
\bBOXRT, \evo, \RESTest, \RestCT, \RESTler, \RestTestGen and \Schemathesis.
Here in this section they are briefly described, in alphabetic order.
For the tool comparisons, we used their latest released versions, as per 
23rd of May  2022.

To the best of our knowledge, this selection represents the current state-of-the-art in fuzzing RESTful APIs,
as those are the most used, most popular (e.g., number of stars and download stats from GitHub) and cited work in the literature. 
For example, some tools in the literature like QuickRest~\cite{karlsson2020QuickREST} do not seem available online,
whereas others like ApiTester~\cite{eddouibi2018automatic} have no documentation and have not been updated in years~\cite{ApiTester}.
The recent tool Morest~\cite{liu2022icse} is not open-source, and its replication package seems that provide no executable nor code to replicate its experiments\footnote{https://anonymous.4open.science/r/morest-rest-8CAE/ currently gives a ``The repository is expired'' error},
 only the data analysis.
So, no comparison seems possible with these other tools.  
In addition, there exist some open-source fuzzers made by practitioners in industry, such as APIFuzzer~\cite{apifuzzer}, Dredd~\cite{dredd} and Tcases~\cite{tcases}.
However, since these fuzzers do not appear in the scientific literature (i.e., they do not have scientific articles describing the techniques they use to generate test cases), and considering that in a previous study~\cite{kim2022arxiv} these tools achieve worse results compared to the other fuzzers, we do not include them in this study.

All these tools require a OpenAPI/Swagger~\cite{Swagger} schema to operate.
Based on such a schema, these tools send syntactically valid HTTP requests, using different strategies (e.g., random and search-based) to choose how to create the input data (e.g., query parameters and JSON body payloads).

\bBOXRT~\cite{laranjeiro2021black} aims at black-box robustness testing of RESTful APIs.
The tool is written in Java, and its source code is available online on the authors' institute pages~\cite{bBOXRT}, since 2020.
We could not find any released version, so we used the latest version on its Git master branch (commit 7c894247).

\evo~\cite{arcuri2017restful,arcuri2018evomaster,arcuri2019restful,arcuri2020sql,arcuri2020blackbox,
arcuri2021evomaster,zhang2021resource,zhang2021adaptive,arcuri2021enhancing}
is a search-based tool that uses evolutionary computation to generate test cases for RESTful APIs.
It supports both black-box and white-box testing~\cite{arcuri2020blackbox} (but this latter only for programs running on the JVM and JavaScript/TypeScript).
The tool is written in Kotlin, and open-sourced on GitHub~\cite{EvoMaster}, since 2016.
For this study, we used version 1.5.0.
We are the authors of this tool.

\RESTest~\cite{restest2020,martinLopez2021Restest,martin2021specification,mirabella2021deep,valenzuela2022arte}   is black-box testing tool, open-sourced on GitHub~\cite{RESTest}, since 2018.
One of the main features of this tool is the ability to analyze inter-dependencies among input parameters.
The tool is written in Java.
We used its 1.2.0 version.

\RestCT~\cite{wu2022icse} uses Combinatorial Testing for black-box testing of RESTful APIs.
It is open-sourced on GitHub~\cite{RestCT}, since 2022.
The tool is written in Python.
We used its 1.0 version.

\RESTler~\cite{restlerICSE2019,atlidakis2020checking,godefroid2020differential,Godefroid2020Restlerdata} is a black-box fuzzer for RESTful APIs.
It is open-sourced on GitHub~\cite{RESTler}, since 2020.
The tool is written in Python and F\#.
It does not have any published release, although its Git repository has tagged commits.
We use the version of the repository with the latest tag \emph{v8.5.0}.

\RestTestGen~\cite{viglianisi2020resttestgen,corradiniautomated2022} is another black-box fuzzer for RESTful APIs.
It is written in Java, but the original versions used in~\cite{viglianisi2020resttestgen} are not open-source.
Since late 2021, a new rewrite of the tool as open-source is available on Github~\cite{RestTestGenV2}.
Then, in this study, we use a version with the latest tag \emph{v22.01}.

\Schemathesis~\cite{hatfield2022deriving} is a black-box fuzzer which employs property-based testing techniques~\cite{maciver2019hypothesis}, whose development started in 2019.
The fuzzer is capable of deriving structure and semantics of SUTs based on their API schemas, such as OpenAPI~\cite{Swagger}.
It is written in Python, and can be installed with \texttt{pip}.
In this study, we use  the version~\emph{3.15.2}~\cite{Schemathesis}.

%%%%%%%%%%%%%%%%%%%%%%%%%%%%%%%%%%%%%%%%%%%%%%%%%%%%%%%%%%%%%%%%%%%%%%%%%%%%
\section{Related Work}
\label{sec:related}

To achieve better results (e.g., higher code coverage and fault detection), we need to understand what are the limitations of current testing techniques.
Throughout the years, different studies have been carried out to provide this insight, like for example studying the limitations of Dynamic Symbolic Execution (DSE)~\cite{Xiao2011,xiao2013characteristic} and Search-Based Software Testing (SBST) for unit testing~\cite{Arc09c,HaM10,albunian2020causes}.
However, to the best of our knowledge, no work has been carried out to study what are the limitations of fuzzing RESTful APIs (most work is for black-box testing, where achieved code coverage is usually not reported, and no analysis of what was not covered is carried out).

Regarding tool comparisons for fuzzing RESTful APIs, there has been some work in the literature.
For example, the authors of \RestTestGen compared four black-box tools (\RestTestGen, \RESTler, \bBOXRT and \RESTest) on 14 APIs~\cite{corradini2021empirical}.
These APIs are written in different languages (e.g., Java, C\#, PHP, JavaScript and Go), with the largest having up to $24 044$ lines of code. 
However, no code coverage was reported. 
From this comparison, \RestTestGen seems the most effective tool, whereas \RESTler is the most robust (i.e., could be used on more APIs without crashing).

The authors of \RESTest and \evo compared the two tools~\cite{martin2021black}, aiming at studying the tradeoffs between black-box and white-box testing.
For example, they studied the impact on performance of using ``custom generators'', with test data provided by the users. 

The authors of \RestCT, when they introduced their tool~\cite{wu2022icse}, compared it with \RESTler on 11 APIs coming from two projects (i.e., GitLab and Bing Maps), showing better results for \RestCT.

The authors of Morest~\cite{liu2022icse} compared it against with black-box \evo, \RestTestGen and \Restler, on 6 APIs.
They claimed Morest gives the best results.
However, as Morest is not available for comparisons (in contrast to \evo, \RestTestGen and \Restler), such claims cannot be independently verified. 

In previous work, we compared \evo's black-box and white-box mode on 8 SUTs~\cite{arcuri2020blackbox}, 7 open-source and 1 industrial, showing better results for its white-box mode.
The SUTs used in this paper are a super-set of the open-source SUTs used in~\cite{arcuri2020blackbox} (i.e., for the experiments on the JVM we used the same SUTs plus another~5).

In parallel, at the same time of the first arXiv version of this paper~\cite{zhang2022open},
Kim et al.~\cite{kim2022arxiv} made a comparison of tools for fuzzing RESTful APIs.
A total of different 9 tools (the same as here, minus the recent \RestCT, but plus the aforementioned APIFuzzer~\cite{apifuzzer}, Dredd~\cite{dredd} and Tcases~\cite{tcases}) were compared on 20 APIs running on the JVM (where half of them are the same as here, which come from our own EMB~\cite{EMB} repository of APIs we use to experiment with \evo). 
To better differentiate from this existing work that was done in parallel,  to better generalize our results, here we include as well 6 JavaScript/TypeScript APIs running on NodeJS, as well one industrial API coming from one of our industrial partners. 
Furthermore, where the focus of~\cite{kim2022arxiv} seems to be the comparison of tools, our focus is on the in-depth analysis of the \emph{open problems} in fuzzing RESTful APIs (Section~\ref{sec:problems}).
The comparison of tools is only a mean to identify the one that gives the best results, as the tests generated by such tool are the starting point of the in-depth analyses. 
In both studies~\cite{zhang2022open,kim2022arxiv} black-box \evo gives the best results, closely followed by \Schemathesis, and white-box testing gives better results than black-box. 
Kim et al.~\cite{kim2022arxiv} do not seem to be authors of any of the compared tools.
On the one hand, their comparison is therefore unbiased. 
This is different from our case, as we are the authors of \evo, which turned out to be the best in all these comparisons.
One should always be a bit skeptical of studies where the tool of the authors gives the best results, especially if such results are not possible to be replicated by third-parties. 
This is one of the main reasons why \evo is open-source, with all its experiment scripts stored in its Git repository, automatically updated on Zenodo  for long term storage at each new release (e.g., version $1.5.0$~\cite{andrea_arcuri_2022_6651631}), as well preparing all the SUTs for experiments in a single repository (i.e., EMB~\cite{EMB}).
On the other hand, being the authors of the most performant tool gives us a unique insight when analyzing and discussing the current open problems in fuzzing RESTful APIs (Section~\ref{sec:problems}).

%%%%%%%%%%%%%%%%%%%%%%%%%%%%%%%%%%%%%%%%%%%%%%%%%%%%%%%%%%%%%%%%%%%%%%%%%%%%
\section{Empirical Analysis}
\label{sec:empirical}

In this paper, we aim at answering the following research questions:
\begin{itemize}
\item {\bf RQ1:} How do the seven compared black-box fuzzers fare in terms of line coverage?
\item {\bf RQ2:} What line coverage and fault detection results are obtained with white-box fuzzing?
\item {\bf RQ3:} What are the main open problems currently hindering the results?
\end{itemize}

To answer these research questions, we carried out 2 different sets of experiments: first for black-box
testing ({\bf RQ1} in Section~\ref{sub:bb}), and then for white-box testing ({\bf RQ2} in Section~\ref{sub:wb}), both on the same case study (describe in Section~\ref{sub:casestudy}).
From the results of these experiments, the in-depth analysis of the results for {\bf RQ3} follows in Section~\ref{sec:problems}.

%----------------------------------------------------------
\subsection{Case Study}
\label{sub:casestudy}

\begin{table}[ht!]
\centering
\caption{\label{tab:sut}
Statistics on 18 RESTful APIs from EMB~\cite{EMB} and one industrial API.}	
\vspace{-1\baselineskip}
\input{generated_files/sut_info.tex}
\begin{spacing}{0.8}
	\raggedright \footnotesize Note that, regarding \textit{Total(X,Y)}, \textit{Total} represents the total of the statistic on all  case studies, whereas \textit{X} represents the total on just the JavaScript/TypeScript APIs, and \textit{Y} represents the total on the JVM APIs.  
\end{spacing}
\end{table}

To carry out experiments in this paper, we used a collection of 19 RESTful APIs. 
As we need to measure code coverage and analyze the source code to check which parts are not covered, 
we needed open-source APIs that we could run on a local machine. 
Furthermore, to simplify the collection of code coverage results, it is easier to use APIs written in the same programming language (e.g., Java), or at least use not too many different languages, as each programming language would need to configure its own code-coverage tool to analyze the test results. 
Considering that we wanted to do comparisons also with white-box testing, which currently only \evo supports, and that requires some manual configurations (e.g., to set up bytecode instrumentation for the white-box heuristics), we decided to use the same case study that we  maintain for \evo.
In particular, we maintain a repository of RESTful APIs called EMB~\cite{EMB}, which is stored as well on Zenodo~\cite{andrea_arcuri_2022_6106830}.
Table~\ref{tab:sut} shows the statistics of these 19 APIs.
Note that one of these APIs is coming from one of our industrial partners, which of course we are not allowed to store on EMB. 

These 19 APIs  are written in four different languages: Java, Kotlin, JavaScript and TypeScript.
They run on 2 different environments/runtimes: the JVM and NodeJS. 
For each SUT, we report the number of total source files (i.e., ending in either \texttt{.java}, \texttt{.kt}, \texttt{.js} or \texttt{.ts}), and their number of lines (LOCs).
As these include as well import statements, empty lines, comments and tests, for the APIs running on the JVM we also report the number of actual line targets for code coverage (measured with the tool JaCoCo~\cite{JaCoCo}).

For the APIs running on the NodeJS, the code coverage is measured with the tool c8, which uses native V8 coverage.
By default, the tool c8 will count code coverage only for the files which are loaded by the engine~\cite{C8}.
For instance, regarding \cyclotron, different LOCs between Files and c8 are due to unreachable files (i.e.,\texttt{ api.analyticselasticsearch.js}, \texttt{api.analytics.js}, \texttt{api.statistics-elasticsearch.js}).
However, all of the files will be only reached by manually modifying a configuration in \texttt{config.js}, i.e., the default value is \texttt{false}.
Therefore, we report the number of line targets measured by c8 that could be loaded with the default SUT settings.

%----------------------------------------------------------
\subsection{RQ1: Black-Box Testing Experiments}
\label{sub:bb}

For each SUT, we created Bash scripts to start them, including any needed dependency (e.g., as \emph{ocvn-rest} uses a MongoDB database, this is automatically started with Docker).
Each SUT is started with either JaCoCo (for JVM) or c8 (for NodeJS) instrumentation, to be able to collect code coverage results at the end of each tool execution. 

Each of the seven compared tools was run on each of the 19 SUTs, repeated for 10 times to keep into account the randomness of these tools.
This results in a total of $7 \times 19 \times 10 = 1330$ Bash scripts.
Each script starts a SUT as a background process, and then one of the tools.
Each script runs the SUT on a different TCP port, to enable running any of these scripts in parallel on the same machine. 

The code coverage is computed based on all the HTTP calls done during the fuzzing process, and not on the output of the generated test files (if any).
This was done for several reasons: not all tools generate test cases in JUnit on JavaScript format, the generated tests might not compile (i.e., bugs in the tools), and setting up the compilation of the tests and running them for collecting coverage would be quite challenging to automate (as each tool behaves differently).
This also means that, if a tool crashes, we are still measuring what code coverage it achieves.
If a tool crashes immediately at startup (e.g., due to failures in parsing the OpenAPI/Swagger schemas), we are still measuring the code coverage achieved by the booting up of the SUT.

In each script, each tool was run for 1 hour, for a total of 1330 hours, i.e., 55.4 days of computation efforts.
The experiments were run on a Windows 10 machine with 
a processor	Intel(R) Xeon(R), 2.40GHz, 24 Cores with 192G of RAM. 
Note: the choice of the runtime for each experiment might impact the results of the comparisons.
The choice of 1 hour is technically arbitrary, but based on what practitioners might want to use these fuzzers in practice, and also not too long to make running all these experiments unviable in reasonable time. 

Regarding the selected seven fuzzers, there exist fuzzers which do not provide an option to configure a global time budget to terminate fuzzing (e.g., \Schemathesis~\cite{Schemathesis_cli} and \RestTest~\cite{RESTest}).
Also, although some fuzzers provide the option of a timeout (e.g., \RestTestGenVT), they might terminate much earlier than the specified timeout value~\cite{corradiniautomated2022}.
To make the comparison of the fuzzers more fair by applying the same time budget, for each fuzzer we run it in a loop with 1 hour timeout (i.e., if the fuzzer runs out of time, it would be terminated, and if the fuzzer completes but there is still some time remaining, it would be re-started to generate more tests).  
Thus, all coverage we collected are based on 1 hour time budget for all the fuzzers.

To compare these tools, we use the line coverage reported by JaCoCo and c8 as metric.
Another important metric would be fault detection.
However, how to compute fault detection in unbiased way, applicable for all compared tools, is far from trivial. 
Each tool can self-report how many faults they find, but how such fault numbers are calculated might be very different from tool to tool, making any comparison nearly meaningless.
Manually checking (possibly tens of) thousands of test cases is  not a viable option either.
Therefore, for this type of experiments line coverage was the most suitable metric for the comparisons. 
We are still going to compare fault detection, but for the white-box experiments (Section~\ref{sub:wb}), as we use the same tool (i.e., \evo).

Some authors like in~\cite{kim2022arxiv} use the number of unique exception stack traces in the SUT's logs as a proxy of detected faults.
On the one hand, a tool that leads the SUT to throw more exceptions could be considered better. 
On the other hand, using such metric as proxy for fault detection has many shortcomings, like for example: (1) not all exceptions reported in the logs are related to faults;
(2) not all SUTs actual print any logs by default (e.g., this is the case for several SUTs in our study);
(3) crashes (which lead to responses with HTTP status code 500) are only one type of faults in RESTful APIs detected by these fuzzers~\cite{marculescu2022faults}, where for example all faults related to response mismatches with the API schema would leave no trace in the logs.
For all these reasons, we did not do this kind of analysis on the logs, as we do not think they provide much more info compared to line coverage.
Especially considering that the infrastructure to do such analysis would need to be implemented, which might not be a trivial task.

Regarding experiment setup, as a black-box testing tool for fuzzing REST APIs, all tools are required to configure where the schema of the API is located.
In all these SUTs used in our case study, the schemas are provided by the SUTs directly via an HTTP endpoint.  
But, we found that most of the tools do not support to fetch the schema with a given URL, such as \url{http://localhost:8080/v2/api-docs}.
To conduct the experiments with these tools, after the SUT starts, we developed a Bash script which manages to fetch the schema and download it to the local file system, and then configure a file-path  for the tools to specify where the schema can be accessed.

Regarding additional setups to execute the tools, \evo, \RestCT
and \Schemathesis 
 were the simplest to configure, as they require only one setup step, as all of their options can be specified with  command-line arguments.
However, \RestCT currently does not work directly on Windows~\cite{RestCT}.
So, for these experiments, we simply ran it via the Windows Subsystem for Linux (WSL). 
\RestTestGenVT requires having a JSON file (i.e., \texttt{rtg\_config.json}) to configure the tool with available options~\cite{RestTestGenV2}.
\Restler 
requires multiple setup steps, e.g., \Restler needs to generate grammar files first, and then employ such grammars for fuzzing. 
However, with its online available documentation, we could write a Python script about how to run the tool.
For \RestTest, it requires a pre-step to generate a test configuration in order to employ the tool, and such generation could be performed automatically by a utility \texttt{CreateTestConf} provided by the tool.
In order to use \bBOXRT, a Java class file is required to load and set up the API specification. 
At the time of writing this paper, there does not exist specific documentation about how to specify such Java class.
However,  in its open-source repository, there exist many examples that helped us to create these Java classes for the SUTs in our case study.
Note that, for these  experiments in this paper, all the above setups were performed automatically with our Bash scripts.

The first time we ran the experiments, we could collect results only for \evo, \RESTler and \Schemathesis.
All the other tools failed to make any HTTP calls.
This was due for example to mismatched schema format, or missing/misconfigured info in the schemas. 
More specifically, \bBOXRT only allows a schema with YAML format.
In the SUTs used in this study, there is only one specified with YAML (i.e., \restcountries) out of the 19 schemas (the remaining ones use JSON). 
\RestTestGenVT only accepts a schema with OpenAPI v3 in a JSON format~\cite{RestTestGenV2}.
There are only two (i.e., \cwa and \spacexapi) out of the 19 SUTs which expose OpenAPI v3 in the JSON format.  
In addition,
\RestTest and \RestTestGenVT need the protocol info (e.g., http or https with \texttt{servers}/\texttt{schemes} tag) in the OpenAPI/Swagger schema. 
But since the \texttt{servers}/\texttt{schemes} tag is not mandatory, such info might not be always available in the schema.
For example, seven (i.e., \cyclotron, \diseasesh, \jsrestncs, \jsrestscs, \realworldapp, \featuresservice and \restcountries) out of the 19 SUTs have such protocol info specified in their schemas. 
In addition, to create HTTP requests, \RestCT, \RestTest and \RestTestGenVT require info specified in \texttt{host} (for schema version 2) and \texttt{servers} (for schema version 3), but such info (typically  related to TCP port numbers) might not be fully correct (e.g., the host and TCP port might refer to the production settings of the API, and not reflecting when the API is running on the local host on an ephemeral port for testing purposes).
Those three tools do not seem to provide ways to override such info.
For instance, in 19 SUTs 
, 10 SUTs (i.e., \cyclotron, \diseasesh, \jsrestncs, \jsrestscs, \realworldapp, \spacexapi, \cwa, \featuresservice, \languagetool and \restcountries) are specified with a hard-coded TCP port, and in one SUT (i.e., \scoutapi) the TCP port is unspecified.
To avoid these issues in accessing the SUTs, we developed a utility using \texttt{swagger-parser} library that facilitates converting formats of schemas between JSON and YAML (only applied for \bBOXRT and \RestTestGenVT), 
converting OpenAPI v2 to OpenAPI v3 (only applied for \RestTestGenVT), adding missing \texttt{schemes} info, and correcting/adding \texttt{host} and \texttt{servers} info in the schemas.

Once these changes in the schemas were applied, we repeated the experiments, to collect data from all the seven tools.
Ideally, these issues should be handled directly by the tools.
But, as they are rather minor and only required changes in the OpenAPI/Swagger schemas, we decided to address them, to be able to collect data from all the seven tools and not just from three of them.

In addition, we needed to configure authentication info for five APIs, namely \proxyprint, \scoutapi, \ocvnrest, \realworldapp and \spacexapi.
For \proxyprint, \scoutapi and \spacexapi, they need \emph{static} tokens sent via HTTP headers.
This was easy to setup in \RestCT, \evo and \Schemathesis, just by calling these tools with a
\texttt{--header} input parameter.
\RESTest required to rewrite the test configuration file by appending authentication configuration. 
\RESTler and \RestTestGenVT 
required to write some script configurations.
\bBOXRT has no documentation to setup authentication, but we managed to setup it up by studying the examples it provides in its repository.

Regarding \ocvnrest and \realworldapp, for authentication, it requires to make a \texttt{POST} call to a form-login endpoint, and then use the received cookie in all following HTTP requests.
Out of the seven compared tools, it seems 
\RESTler, 
\bBOXRT,
\Schemathesis,  
and \RestTestGenVT
 could directly support this kind of authentication by setting up it with an executable script. 
Given the provided documentation, we did not manage to 
configure it,
as it requires to write different scripts for different fuzzers 
to manually make such HTTP login calls then handle responses. 
Technically, by writing manual scripts, it could be possible to use  \evo, \RestCT and \RestTest as well, by passing the obtained cookie with \texttt{--header} option or the test configuration file.
As doing all this was rather cumbersome, and considering that for this API the authentication is needed only for \emph{admin} endpoints, we decided to do not spend significant time in trying to setup this kind \emph{dynamic} authentication tokens.

Table~\ref{tab:bb} shows the results of these experiments.
For each tool, we report the average line coverage, as well as the min and max values out of the 10 runs. 
Each tool is then ranked (1 to 7) based on their average performance on each SUT (where 1 is the best rank).

From these results we can infer few interesting observations.
First, regarding the ranking, \evo seems the best black-box fuzzer (best in 11 out of 19 SUTs, with an average coverage of 56.8\%,),
closely followed by \Schemathesis (best in 7 SUTs, with an average coverage of 54.5\%).
Then, the remaining tools can be divided in 2 groups:  \bBOXRT and \RestTestGenVT with similar coverage 41.6-45.4\%,
and then  \RESTler, \RESTest and  \RestCT with similar coverage 33.6-35.4\%. 
These results confirm a previous study~\cite{corradini2021empirical} showing that  \RestTestGenVT gives better results than \RESTler and \RESTest, as well as  \RestCT being better than \RESTler~\cite{wu2022icse} (although in this case the difference in average coverage is small, only 0.8\%).
Compared to the analyses in~\cite{kim2022arxiv}, interestingly the ranking of the tools is exactly the same (recall that, out of the combined 29 APIs between our and their study, only 10 APIs used in these empirical studies are the same).

On all APIs but 1, either \evo or \Schemathesis gives the best results.
The exception is the industrial API, where 5 tools achieve the same coverage of 8.2\%, on all their 10 runs. 
We will discuss this interesting case in more details in Section~\ref{sec:problems}.  
In 12 APIs, either \evo is the best followed by \Schemathesis, or the other way round. 
There is no single API in which \evo and \Schemathesis were not at least the third best.

The other 7 APIs (including the industrial one) show some interesting behavior for the other 5 tools.
For example, for \emph{js-rest-scs}, \RestTestGen gives the second-best results,
with an average 86.4\%,  compared to the 86.1\% of \Schemathesis.
The interesting aspect here is that, out of the 10 runs, \RestTestGen has worse minimum coverage (85.4\% vs.~85.9\%) and worse maximum coverage  (87.1\% vs.~87.4\%), although the average is higher (86.4\% vs.~86.1\%). This can happen when randomized algorithms are used. 
In \emph{gestaohospital-rest}, \RestTestGen and \Schemathesis have similar performance (i.e., 57.2\% and 58.7\%), whereas \evo is quite behind (i.e., 50.8\%).
Similarly, the performance of \RestTestGen and \Schemathesis are very similar on \emph{rest-scs} (65.3\% and 64.8\%) and \restcountries (75.4\% and 73.9\%), where \evo is better only by a small amount (66.9\% on \restscs and 76.1\% on \restcountries). 
In \emph{languagetool}, \RestTest is better than \Schemathesis, but the difference is minimal (only 0.3\%).
In \emph{rest-ncs}, there is large gap in performance between \evo (64.5\%) and \Schemathesis (94.1\%), where the second best results are given by \RestCT (85.5\%).
Finally, on \emph{scout-api}, \Restler  is better than \Schemathesis (26.5\% vs.~23.0\%), although it is way behind \evo (36.7\%).

Another interesting observation is that there is quite a bit variability in the results of these fuzzers, as they use randomized algorithms to generate test cases.
Let us highlight some of the most extreme cases in Table~\ref{tab:bb}, for \evo and \Schemathesis.
On \emph{languagetool}, out of the 10 runs, \evo has a gap of 9.1\% between the best (35.1\%) and worst (26.0\%) runs. 
On \emph{scout-api}, the gap is 8.3\%.
For \Schemathesis, the gap on \emph{features-service} is 10.5\%, and 6.4\% on \emph{gestaohospital-rest}.
This is yet another reminder of the peculiar nature of randomized algorithms, and the importance of how to properly analyze them.
For example, doing comparisons based on a single run is unwise.

Statistical tests~\cite{Hitchhiker14} are needed when claiming with high confidence that one algorithm/tool is better than another one. 
In this particular case, we compare \evo's performance with all the other tools, one at a time on each SUT (so $6 \times 19=114$ comparisons), and report the $p$-values of the Mann-Whitney-Wilcoxon U-Test in Table~\ref{tab:pvalues}.
Apart from very few cases, the large majority of comparisons are statistically significant at level $\alpha=0.05$.  
Often, 10 repetitions might not be enough to detect statistically significant differences, and higher repetition values like 30 and 100 are recommended in the literature~\cite{Hitchhiker14}.
However, here the performance gaps are large enough that 10 repetitions were more than enough in most cases.

When looking at the obtained coverage values, all these tools achieve at least a 30\% coverage on average. 
Only two of them (i.e., \evo and \Schemathesis) achieve more than 50\%.
But no tool goes above 60\% coverage.
This means that, although these tools might be useful for practitioners, there are still several research challenges that need to be addressed (we will go in more details on this in Section~\ref{sec:problems}).
However, what level of coverage can be reasonably expected from black-box tools (which have no info on the source code of the SUTs) is a question that is hard to answer.

\begin{result}
	{\bf RQ1:} all compared tools achieve at least 30\% line coverage on average, but none goes above 60\%. Of the 7 compared tools, \evo  seems the one giving the best results, closely followed by \Schemathesis. 
\end{result}

\begin{landscape}

	\begin{table*}%[ht!]
		\centering
		\caption{\label{tab:bb}  
			Experiment results of the seven black-box fuzzers on the 19 RESTful APIs.
			For each tool we report the average line coverage, as well as its [min,max] values out of the 10 runs.
			Each tool also has a \emph{(rank)} based on its performance on each SUT.
			For each SUT, the results of the best tools (e.g., rank (1)) are in bold.   
		}	
		\vspace{-1\baselineskip}
		\resizebox{.99\linewidth}{!}{
			\input{generated_files/tableBB_1h.tex}
		}
	\end{table*}
	
\end{landscape}

\begin{table}[ht!]
	\centering
\caption{\label{tab:pvalues}  
$p$-values of the Mann-Whitney-Wilcoxon U-Test of \evo's results compared to all the other tools.
Values lower than the $\alpha=0.05$ threshold are reported in bold.  
	}	
	\resizebox{.99\textwidth}{!}{
	\input{generated_files/tableUTests_1h.tex}
	}
\end{table}

%----------------------------------------------------------
\subsection{RQ2: White-Box Testing Experiments}
\label{sub:wb}

\begin{table}[ht!]

	\centering
\caption{\label{tab:wb}  
Experiment results for white-box and grey-box testing using \evo on the 19 RESTful APIs.
We report the average line coverage measured with \evo itself (not JaCoCo, nor c8), as well as the number of faults detected in these APIs. When the differences are statically significant at $\alpha=0.05$ level, the effect-sizes $A_{12}$ are reported in bold. 
}
		\input{generated_files/tableComparisonWB_1h.tex}
\end{table}

Out of the seven compared tools, only \evo supports white-box testing.
\evo uses evolutionary computation techniques, where the bytecode of the SUTs is instrumented to compute different kinds of heuristics.
Due to possible conflicts with JaCoCo and c8 instrumentation, and due to the fact that \evo uses its own driver classes (which need to be written manually) to start and stop the instrumented SUTs, this set of experiments was run differently compared to the black-box ones.

We ran \evo on each SUT for 10 repetitions (so, 190 runs), each one for 1 hour (like for the black-box experiments).
However, JaCoCo and c8 are not activated.
After each run, \evo generates statistic files, including information on code coverage and fault detection.
However, as there can be differences on how line coverage is computed between \evo and JaCoCo/c8, it would be difficult to reliably  compare with the results in Table~\ref{tab:bb}.
Therefore, for the comparisons, we ran \evo as well in grey-box mode (for another 190 runs).
This mode generates test cases in exactly the same way as black-box mode, with difference the SUT is instrumented, and coverage metrics are computed at each test execution. 
A further benefit is that, besides code coverage, we can reliably compare fault detection as well, as this metric is computed in exactly the same way (as it is the same tool).
As \evo is the black-box fuzzer that gives the highest code coverage (Section~\ref{sub:bb}), it is not a major validity threat to compare white-box results only with \evo.
These 380 runs added a further 15.8 days of computational effort.

Table~\ref{tab:wb} shows these results.
Few things appear quite clearly.
First, on average, line coverage goes up by 7.5\% (from 45.4\% to 52.9\%).
Even with just 10 runs per API, results are statistically significant in most cases. 

For APIs like \restncs, average coverage can go even higher than 90\%.
For other APIs like \featuresservice improvements are more than 13\% (e.g., from 68.8\% to max 81.1\%).
In the case of \indCS, although achieved coverage is relative low (i.e., less than 20\%), it still more than double (i.e., from 8.4\% to 18.8\%).

Although results are significantly improved compared to black-box testing, for nearly half of these SUTs it was still not possible to achieve more than 50\% line coverage.
Also, there are two interesting cases in which results with white-box testing are actually significantly \emph{worse}, i.e., for \emph{realworld-app} and \emph{ocvn-rest}. 
In the former case, the difference is minimal, just 0.2\%.
In the latter case, though, the difference is substantial, as it is 10.5\% (from 35.3\% down to 24.8\%).
This is not an unusual behavior for search algorithms, as it all depends on the quality of the \emph{fitness function} and the properties of the \emph{search landscape}~\cite{campos2018empirical}.   
If a fitness function gives no gradient to the search algorithm, it can easily get stuck in local optima.
In such cases, a random search can give better results. 
\evo uses several heuristics in its fitness function to try to maximize code coverage, but those do not work for \emph{ocvn-rest}, as we will discuss in more details in Section~\ref{sec:problems}.
However, improving the fitness function in this case can be done (which we will address in future versions of \evo). 

Regarding fault detection, white-box testing leads to detect more faults.
This is not unexpected, as usually there is a strong correlation between code coverage and fault detection~\cite{bohme2022reliability}.
For example, you cannot detect a fault if the code it hides in is never executed.
The interesting aspect here is that the improvement is not much, just 3.4 more faults on average.
In these experiments, random testing can find 36.5 faults on average, and so the relative improvement is less than 10\%.
For the problem domain of fuzzing RESTful APIs (and likely Web APIs in general), this is not surprising~\cite{marculescu2022faults}.
Many of these APIs do crash as soon as an invalid input is provided, instead of returning a proper user error response (e.g., with HTTP status code in the 4xx family).  
And random search is quite good at generating invalid input values. 
So, many faults can be easily found this way with a fuzzer at the very first layer of input validation in the SUT's code, even when the achieved code coverage is relatively low.

\begin{result}
{\bf RQ2:} White-box fuzzing leads to significantly higher  results compared to black-box fuzzing, up to  7.5\% more code coverage and 3.4 more detected faults on average. But, still, for many APIs  coverage results were less than 50\%. 
\end{result}

%%%%%%%%%%%%%%%%%%%%%%%%%%%%%%%%%%%%%%%%%%%%%%%%%%%%%%%%%%%%%%%%%%%%%%%%%%%%
\section{RQ3: Open Problems}
\label{sec:problems}

We ran seven tools/configurations on 19 RESTful APIs, with a budget of 1 hour per experiment.
However, only in one single case it was possible to achieve 100\% line coverage (i.e., \Schemathesis on \emph{js-rest-ncs}, recall Table~\ref{tab:bb}).
In general, it might not be possible to achieve 100\% line coverage, as some code can be unreachable.
This is for example a common  case for constructors in static-method-only classes, and catch blocks for exceptions that cannot be triggered with a test.
However, for several of these SUTs it was not even possible to reach 50\% line coverage.

It is not in the scope of this paper to define what would be a good coverage target to aim for (80\%? 90\%?). 
However, clearly the higher the code coverage the better it would be for practitioners using these fuzzers.
So, to improve those fuzzers, it is of paramount importance to understand what are their current limitations.
To answer this question, we studied in details the logs of those tools in Section~\ref{sub:logs}, and large parts of the source code of the SUTs in Section~\ref{sub:code} (recall that those are more than 280 thousand LOCs, see Table~\ref{tab:sut}).
An in-depth analysis of the current open-problems in fuzzing RESTful APIs is an essential scientific step to get more insight into this problem.
This is needed to be able to design novel, more effective techniques.

To be useful for researchers, these analyses need to be ``low-level'', with concrete discussions about the source code of these APIs.  
No general theories can be derived without first looking at and analyzing the single instances of a scientific/engineering phenomenon. 
As the following software engineering discussions might be considered as ``dull'' for some readers, or if the reader is not currently active in the development of a fuzzer, we suggest to jump  directly to our summarizing discussions in Section~\ref{sub:discussion}.

It is important to stress out that the goal of these analyses is to identify general problems, or instances of problems that are likely going to be present in other SUTs as well. 
Designing novel techniques that just \emph{overfit} for a specific benchmark is of little to no use.
Ultimately, what is important will be the results that the practitioners can obtain when using these fuzzers on their APIs.

%-------------------------------------------------------------------------
\subsection{Analysis Of The Logs}
\label{sub:logs}

From the tool logs, at least four common issues are worth to discuss.
First, OpenAPI/Swagger schemas might have some \emph{errors} (e.g., this is the case for \cyclotron, \diseasesh, \cwa, \emph{features-service} \emph{proxyprint}, and \ocvnrest).
This might happen when schemas are manually written, as well as when they are automatically derived from the code with some tools/libraries (as those tools might have faults).
In these cases, most fuzzers just crash, without making any HTTP call or generating any test case.
It is important to warn the users of these issues with their schema, but likely fuzzers should be more \emph{robust} and not crash, e.g., endpoints with schema issues could be simply skipped.

The second issue can be seen in \emph{languagetool}.
Most fuzzers for RESTful APIs support HTTP body payloads only in JSON format.
But, in HTTP, any kind of payload type can be sent.
JSON is the most common format for RESTful APIs~\cite{neumann2018analysis}, 
but there are others as well like XML and the \texttt{application/x-www-form-urlencoded} used in \emph{languagetool}.  
From the results in Table~\ref{tab:bb}, it looks like only \evo supports this format.
On this API, \evo achieves between 26\% and 35.1\% code coverage, whereas no other fuzzer achieves more than 2.5\%. 
 
The third issue is specific to \emph{scout-api}, which displays a special case of the JSON payloads. 
Most tools assume a JSON payload to be a \emph{tree}: e.g., a root object \texttt{A} that can have fields that are object themselves (e.g., \texttt{A.B}), and so on recursively (e.g., \texttt{A.B.C.D} and \texttt{A.F.C}), in a tree-like structure. 
However, an OpenAPI/Swagger schema can define objects that are \emph{graphs}, as it is the case for \emph{scout-api}.
For example, an object \texttt{A} can have a field of type \texttt{B}, but \texttt{B} itself can have a field of type \texttt{A}. 
This recursive relation creates a graph, that needs to be handled carefully when instantiating \texttt{A} (e.g., optional field entries can be skipped to avoid an infinite recursion, which otherwise would lead the fuzzers to crash).

The fourth issue is related to the execution of HTTP requests towards the SUTs.
To test a REST API, 
the fuzzers build  HTTP requests based on the schema,  and then use different HTTP libraries to send the requests toward the SUT, e.g., \texttt{JerseyClient} is used in \evo.
By analyzing the logs, we found that for several case studies, some fuzzers seem to not have problems in parsing schemas, but they failed to execute the  HTTP requests.
For instance, \texttt{javax.net.ssl.SSLException: Unsupported or unrecognized SSL message} was thrown when \RestTestGenVT processed \realworldapp with \texttt{OkHttpClient}.
For \jsrestncs, \jsrestscs and \indCS, we found that \texttt{404 Not Found} responses were always returned when fuzzing them with \Restler v8.5.0.
By checking the logs obtained by \Restler, we found that it might be due to a problem in generating the right URLs for making the HTTP requests.
For the SUT whose \texttt{basePath} is \texttt{/}, \Restler seems to generate 
double slash (i.e., \texttt{//}) in the URL of the requests.
However, whether to accept the double slash to match a real path depends on the SUTs, i.e., \jsrestncs, \jsrestscs and \indCS do not allow it.
In Figure~\ref{fig:restler_ncs}, we provide the logs obtained by \Restler, representing the processed requests which contain the double slash and responses returned by \jsrestncs (Figure~\ref{fig:restler_jsNCS}) and \restncs (Figure~\ref{fig:restler_jvmNCS}).

\begin{figure}
	\centering
	\begin{subfigure}[h]{0.95\textwidth}
		\centering
\begin{lstlisting}
2022-06-12 12:30:28.112: Sending: 'GET //api/fisher/1/1/1.23 HTTP/1.1\r\nAccept: application/json\r\nHost: localhost:25900\r\nContent-Length: 0\r\nUser-Agent: restler/8.5.0\r\n\r\n'

2022-06-12 12:30:28.126: Received: 'HTTP/1.1 404 Not Found\r\nX-Powered-By: Express\r\nContent-Security-Policy: default-src \'none\'\r\nX-Content-Type-Options: nosniff\r\nContent-Type: text/html; charset=utf-8\r\nContent-Length: 159\r\nDate: Sun, 12 Jun 2022 10:30:28 GMT\r\nConnection: keep-alive\r\nKeep-Alive: timeout=5\r\n\r\n<!DOCTYPE html>\n<html lang="en">\n<head>\n<meta charset="utf-8">\n<title>Error</title>\n</head>\n<body>\n<pre>Cannot GET //api/fisher/1/1/1.23</pre>\n</body>\n</html>\n'\end{lstlisting}
		\caption{\Restler on \jsrestncs built with NodeJS and Express. By default, a path with extra slash will not match the correct endpoint.}
		\label{fig:restler_jsNCS}
	\end{subfigure}
	\hfill
	\begin{subfigure}[h]{0.95\textwidth}
		\centering
\begin{lstlisting}
2022-06-11 22:55:56.651: Sending: 'GET //api/fisher/1/1/1.23 HTTP/1.1\r\nAccept: application/json\r\nHost: localhost:24850\r\nContent-Length: 0\r\nUser-Agent: restler/8.5.0\r\n\r\n'

2022-06-11 22:55:56.697: Received: 'HTTP/1.1 200 \r\nContent-Disposition: inline;filename=f.txt\r\nContent-Type: application/json;charset=UTF-8\r\nTransfer-Encoding: chunked\r\nDate: Sat, 11 Jun 2022 20:55:55 GMT\r\n\r\n38\r\n{"resultAsInt":null,"resultAsDouble":0.5328886540720141}\r\n0\r\n\r\n'\end{lstlisting}
		\caption{\Restler on \restncs built with Spring Boot 2.0.3 whose default path matching strategy allows the double slash. }
		\label{fig:restler_jvmNCS}
	\end{subfigure}
	\caption{Different handling of requests with different techniques}
	\label{fig:restler_ncs}
\end{figure}

%-------------------------------------------------------------------------
\subsection{Analysis Of The Source Code}
\label{sub:code}

For each SUT, we manually ran the best (i.e., highest code coverage) test suite (which are the ones generated by \evo) with code coverage activated.
Then, in an IDE, we manually looked at which branches (e.g., \texttt{if} statements) were reached by the test case execution, but not covered.  
As well as looking at the cases in which the test execution is halted in the middle of a code block due to thrown exceptions.  
This is done to try to understand what are the current issues and challenges that these fuzzers need to overcome to get better results. 
We will discuss here each SUT, in alphabetic order, one at a time.

Note that, when we refer to the code coverage achieved by white-box \evo, we refer to the results in Table~\ref{tab:wb}.
These coverage values are computed with the instrumentation of \evo itself, and they are not exactly the same as what other coverage tools like JaCoCo and c8 would report (as there can be differences on how coverage is computed). 
For example, coverage results reported by c8 might be significantly higher, as those are computed only on the script files that are loaded (based on all the experiments), whereas for \evo all source files were considered. 
In other words, if on each run $i$ a tool cover $X_i$ lines out of $N_i$ total reported, then the coverage percentage $c_i$ is computed as $c_i=X_i/N$, where $N=max(N_i)$. 
Furthermore, a current limitation of \evo for JavaScript is that the recorded coverage does not include the coverage achieved at boot-time, but only from when the search starts.
This was an issue that has been fixed for the JVM, but not for JavaScript yet.  
Therefore, the numbers in Table~\ref{tab:wb} are not directly comparable with the numbers in Table~\ref{tab:bb}.

\subsubsection{catwatch}
With an average line coverage of 50\%, this can be considered a non-trivial API. 
The main issue is that this SUT makes a call to an external service (more specifically, to the GitHub APIs to fetch project info).
But such call seems to get stuck for a long time (and possibly timeout), which leads to none of the code parsing the responses (and do different kind of analyses) being executed. 
Calling external services is a problem for fuzzers.
External services can go down or change at any time.
They can return different data at each call, making assertions in the generated tests become flaky. 
Although testing with the actual external services is useful and should be done, the generated tests would likely not be suitable for regression testing. 
This is a known problem in industry, and there are different solutions to address it, like \emph{mocking} the external services (e.g., in the JVM ecosystem, WireMock~\cite{WireMock} is a popular library to do that).
Enhancing fuzzers to deal with mocked services (e.g., to setup the mock data they should return) is going to be an important venue of future research.

\subsubsection{cwa-verification}

\begin{figure}
\begin{lstlisting}[language=java,basicstyle=\footnotesize]
@PostMapping(value = TAN_ROUTE,
    consumes = MediaType.APPLICATION_JSON_VALUE,
    produces = MediaType.APPLICATION_JSON_VALUE
)
public DeferredResult<ResponseEntity<Tan>> generateTan(
           @Valid @RequestBody RegistrationToken registrationToken,
           @RequestHeader(value = "cwa-fake", required = false) String fake) {
    if ((fake != null) && (fake.equals("1"))) {
      return fakeRequestService.generateTan(registrationToken);
    }
    StopWatch stopWatch = new StopWatch();
    stopWatch.start();
    Optional<VerificationAppSession> actual
      = appSessionService.getAppSessionByToken(registrationToken.getRegistrationToken());
    if (actual.isPresent()) {                            
\end{lstlisting}
\caption{\label{fig:cwa:external}
A function snippet from the \texttt{ExternalTanController} class in the API \cwa.
}
\end{figure}

On this API, achieved coverage is less than 50\% (i.e., 46.9\%).
There are two interesting cases to discuss here, which have major impact on the achieved coverage.

\sloppy
First, Figure~\ref{fig:cwa:external} shows a snippet of code for one of the endpoint handlers (i.e., \texttt{ExternalTanController}, but the same issue happens as well in \texttt{ExternalTestStateController} and \texttt{InternalTanController}).
Here, the condition \texttt{actual.isPresent()} is never satisfied, which leads to miss large part of the code.
A token is given as input as part of the body payload, and then a record matching such token in the database is searched for.
However, none is found, and \evo is unable to create it directly.
In theory, such case should be trivial to handle with SQL support~\cite{arcuri2020handling}, but it was not the case.
The reason is the peculiar properties of such token.
Such token has the given constraint in the OpenAPI schema:

\begin{lstlisting}[language=java,basicstyle=\footnotesize]
"registrationToken": {
 "pattern": "^[a-f0-9]{8}-[a-f0-9]{4}-4[a-f0-9]{3}-[89aAbB][a-f0-9]{3}-[a-f0-9]{12}$",
 "type": "string"
}
\end{lstlisting}          

This results in the token having a length of 36 characters. 
\evo has no problem in sampling strings that match such a regular expression, like for example:
\texttt{bfe8b80b-dca1-458d-B312-96ecae446be0}.
However, such constraint is not present in the SQL database, where the column for these tokens is simply declared with:

\begin{lstlisting}[language=java,basicstyle=\footnotesize]
- column:
    name: registration_token_hash
    type: varchar(64)
\end{lstlisting}

When \evo generates data directly into the database, they would be random strings that do not satisfy such regular expression. 
Again, this should not a problem thanks to taint analysis~\cite{arcuri2021enhancing}.
The reason it is not working is due to the length of the string, which is 36 characters.
By default, \evo does not generate random strings with more than 16 characters.
Even if with taint analysis we could inject the right string into the database, currently this does not happen due to 36 being greater than 16. 

Note that having a constraint on the length of strings is essential. 
Sampling unbound random strings with billions of characters would have too many negative side effects. 
The choice of the value 16 is arbitrary: could had been higher, or lower.   
Still, whatever limit is chosen, an SUT could need strings longer than such limit, as it happens in this case for \cwa.

A solution to address this issue is to distinguish between 2 different max-length limits:
an \emph{hard} one that should never be violated (e.g., a constraint in the OpenAPI or SQL schemas),
and a \emph{soft} one (e.g., 16). 
The soft limit would be used when sampling random strings, but could be violated in some specific circumstances (like for example in taint analysis).

\begin{figure}
\begin{lstlisting}[language=java,basicstyle=\footnotesize]
@PostMapping(value = TELE_TAN_ROUTE,
    produces = MediaType.APPLICATION_JSON_VALUE
)
public ResponseEntity<TeleTan> createTeleTan(
    @RequestHeader(JwtService.HEADER_NAME_AUTHORIZATION) @Valid AuthorizationToken authorization,
    @RequestHeader(value = TELE_TAN_TYPE_HEADER, required = false) @Valid TeleTanType teleTanType) {

    List<AuthorizationRole> requiredRoles = new ArrayList<>();

    if (teleTanType == null) {
      teleTanType = TeleTanType.TEST;
      requiredRoles.add(AuthorizationRole.AUTH_C19_HOTLINE);
    } else if (teleTanType == TeleTanType.EVENT) {
      requiredRoles.add(AuthorizationRole.AUTH_C19_HOTLINE_EVENT);
    }
\end{lstlisting}
\caption{\label{fig:cwa:internal}
A function snippet from the \texttt{InternalTanController} class in the API \cwa.
}
\end{figure}

The second interesting case to discuss is present in the endpoint handler \texttt{InternalTanController},
shown in Figure~\ref{fig:cwa:internal}. 
Here, an HTTP header with value \texttt{TELE\_TAN\_TYPE\_HEADER="X-CWA-TELETAN-TYPE"} can be provided as input.
However, such info is missing from the OpenAPI schema.
Therefore, the object \texttt{teleTanType} is always null. 
This is  an example of underspecified schema.

\subsubsection{cyclotron}

On this API, there is not much difference between white-box and grey-box testing.
There are at least 3 major issues that are worth to discuss.

First, a non-negligible amount of code is executed only if some configuration settings are \emph{on}.
But those are \emph{off} by default (like for example \texttt{config.analytics.enable} and \texttt{config.enableAuth} in \texttt{config.js}).
All the code related to these functionality cannot be tested via the RESTful endpoints.

\begin{figure}[h]
\begin{lstlisting}[language=java,basicstyle=\footnotesize]
exports.upsertData = function (req, res) {
    if (req.body == null) {
        return res.status(400).send('Missing data.');
    }

    var upsertData = req.body.data;
    var keys = req.body.keys;

    if (upsertData == null) {
        return res.status(400).send('Missing data.');
    }
     if (keys == null) {
        return res.status(400).send('Missing keys.');
    }
\end{lstlisting}
\caption{\label{fig:cyclotron}
A function snippet from the \texttt{api.data.js} file in the API \cyclotron, 
for the endpoint \texttt{/data/\{key\}/upsert}.
}
\end{figure}
  
\begin{figure}[h]
\begin{lstlisting}[language=java,basicstyle=\footnotesize,escapechar=©]
"/data/{key}/upsert": {
  "post": {
    "summary": "Upserts an object in the Data Bucket Data",
    "description": "Upserts (inserts or updates) an object in the data for a given Data Bucket. The rev property will be incremented.",
    "tags": [
       "Data"
     ],
     "parameters": [{
          "name": "key",
          "in": "path",
          "description": "The Data Bucket key.",
          "required": true,
          "type": "string"
        }, {
          "name": "data",
          "in": "body",
          "description": "An object containing 'keys' and 'data'.", ©\label{line:upsert}©
          "required": true
        }],
\end{lstlisting}
\caption{\label{fig:cyclotron:schema}
Snippet of the OpenAPI schema for the API \cyclotron.
}
\end{figure}

Second, like for several other APIs in this study, the schema here is not fully correct/complete.
For example, Figure~\ref{fig:cyclotron} shows a snippet for the handler of the endpoint 
\texttt{/data/\{key\}/upsert}, where the two fields \texttt{data} and \texttt{keys} 
are read from the input object in the HTTP body payload of the request.
However, the schema  has no formal definition about those fields, as they are just mentioned as a comment (see Line~\ref{line:upsert} in Figure~\ref{fig:cyclotron:schema}).

Third, the API does several accesses to a MongoDB database.
Several endpoints start by retrieving data from the database, like for example
with commands like \texttt{Dashboards.findOne(\{ name: dashboardName \})}.
If the data is missing, no following code is then executed.
But it is hard to get the right id (e.g., \texttt{dashboardName} in this example) by chance.
Although \evo has support for SQL databases (i.e., to analyze all executed queries at runtime), it does not for MongoDB, and not for NodeJS (i.e., SQL support is currently only implemented for the JVM).

\subsubsection{disease-sh-api}

On this API, there is not much difference between \evo (both black-box and white-box) and \Schemathesis. 
After an analysis of the generated tests, practically all achievable coverage is obtained, as the remaining code is not reachable through the REST endpoints in the API.

First, the schema refers only to version \texttt{v3} of the API, but in the source code there is still all the implementation of the endpoints for version \texttt{v2}.
Those endpoints are not executable based on the info in the schema.
Second, large part of the code deals with the fetching of disease records from online databases, and store them in a local Redis instance. 
But such code is executed only via scripts, and not from the RESTful endpoints.
Note that, when the experiments are run, the database was populated with a selection of valid data, as there is no REST endpoint to add it.

\subsubsection{features-service}
\begin{figure}
\begin{lstlisting}[language=java,basicstyle=\footnotesize]
private EvaluationResult addEvaluationsToResult(
         EvaluationResult result, 
         ProductConfiguration configuration, 
         Set<FeatureConstraint> featureConstraints){
 for(FeatureConstraint featureConstraint 
                               : featureConstraints) {
   result = featureConstraint.evaluateConfiguration(
                                  result,configuration);
 }

 if(result.isValid && !result.derivedFeatures.isEmpty()){
   for(String derivedFeature : result.derivedFeatures) {
       configuration.active(derivedFeature);
       result.derivedFeatures.remove(derivedFeature);
       result = this.addEvaluationsToResult(
\end{lstlisting}
\caption{\label{fig:features-service}
A function snippet from the \texttt{ConfigurationEvaluator} class in the API \featuresservice.
}
\end{figure}

With an average line coverage of  81.8\%, \emph{features-service} can be considered one of the easier APIs. 
Quite a bit of code cannot be reached with any generated system test, like for example getters/setters that are never called, or some functions that are only used by the manually written unit tests.
There are still some branches that are not covered, related to properties of data in the database.
In other words, when a \texttt{GET} is fetching some data, some properties are checked one at a time, but, to be able to pass all these checks, a previous \texttt{POST} should had created such data. 
Figure~\ref{fig:features-service} shows one such case, where the code inside \texttt{if} statement does not get executed.
Tracing how database data is impacting the control flow execution, and which operations should be first called to create such data, is a challenge that needs to be addressed.

\subsubsection{gestaohospital-rest}

On this API, there is no much difference between white-box and grey-box \evo (i.e., average coverage 39.5\% vs.~39.4\%, Table~\ref{tab:wb}).
Furthermore, \RestTestGen and \Schemathesis give better results than \evo (Table~\ref{tab:bb}).

One specific property of this API compared to most of the other SUTs in our empirical study is that such API uses a MongoDB database.
In contrast to SQL databases, white-box \evo has no support for NoSQL databases (such as MongoDB). 
As this API does many interactions with the database (e.g., there is a lot of calls like \texttt{service.findByHospitalId(hospital\_id)}), not much coverage is achieved if such accessed data is not present in the database, and the evolutionary search has no gradient to create it. 
In this regard, it seems that \RestTestGen and \Schemathesis do a better job at creating resources with \texttt{POST} commands and use such created data for their following \texttt{GET} requests. 
Still, although quite good, the coverage is at most 62.3\% (measured with JaCoCo for \Schemathesis), which means there still some challenges to overcome.

We could not analyze in details the output of these tools in an IDE for the Java program \gestao (e.g., using a debugger), as \RestTestGen generated only JSON files with the descriptions of the tests (and no JUnit file), and the outputs of \Schemathesis are in YAML for the VCR-Cassette format (which is mainly for Ruby and Python, where the port for Java seems has not been under maintenance for several years).

\subsubsection{ind0}

This is a closed-source API provided by one of our industrial partner.
Therefore, we cannot provide any code example for it, but still we can discuss it at a high level.
This API is a service in a microservice architecture.
Although in terms of size it is not particularly big (recall Table~\ref{tab:sut}), it is the most challenging API in our case study.
No black-box fuzzer achieves more than 8.2\% line coverage, and white-box testing goes only up to 18.8\% (on average).

Out of 20 endpoints, all but one endpoint require string inputs satisfying a complex regular expression, but such info is not present in the OpenAPI schema.
So, a random string is extremely unlikely to satisfy such constraint. 
Furthermore, the constraint is expressed in a \texttt{@} annotation, which is handled by the Spring framework before any of the HTTP handlers are executed. 

Thanks to taint-analysis, \evo can handle these cases~\cite{arcuri2021enhancing}, even when constraints are evaluated in third-party libraries (and not just in the business logic of the SUT).
Still, there is quite a bit of variability in the results, i.e., given the average coverage 18.8\%, the standard variation is 7.0, with coverage going from a minimum of 11.9\% to a maximum of 32.6\% (out of the 10 repeated experiment runs). 
\evo manages to solve these constraints, but not in all runs, as such strings not only get matched with a regular expression, but then are farther checked with other constraints. 
However, as the regular expression is exactly the same for all these 19 endpoints, applied on a URL path variable with the same name, it should be possible to design strategies to re-use such information to speed up the search, instead of resolving the same constraints each time for each endpoint. 
Even with the run with highest coverage 32.6\%, many endpoints are not covered due to this issue.

\subsubsection{js-rest-ncs}

This is an artificial API, using numerical-computation functions (e.g., Triangle Classification~\cite{Arc09}) behind REST endpoints.
As this is a re-implementation in JavaScript of \emph{rest-ncs}, we will defer to Section~\ref{sub:sut:ncs} for the analyses.

One thing to notice though is that the coverage values are higher for a very simple reason: the fetching of the OpenAPI schema is part of business logic of the API, and code coverage is computed for it (as the schema is inside a JavaScript file as a string inside an HTTP endpoint declaration).
On the other hand, for Java, the schema is handled as a JSON resource.

\subsubsection{js-rest-scs}

Similarly to the case of \emph{js-rest-ncs}, \emph{js-rest-scs} is a JavaScript re-implementation of \emph{rest-scs}, which involved string-based computation functions.
We therefore will defer to Section~\ref{sub:sut:scs} for the analyses.

As for \emph{js-rest-ncs}, the fetching of the schema impacts the computed code coverage. 
However, for white-box \evo, coverage is worse for \emph{js-rest-scs} than for \emph{rest-scs}.
The reason is rather simple: the support for white-box testing of JavaScript code in \evo is a much more recent addition~\cite{js2022}, and not all features that are implemented for the JVM (e.g., different forms of taint analysis~\cite{arcuri2021tt}) are currently supported.

\subsubsection{languagetool}
On this API, coverage is up to 39.5\%.
This is the largest and most complex API in our study (recall Table~\ref{tab:sut}).
This is a CPU bound (e.g., no database) application, doing complex text analyses. 
This API has only two endpoints: \texttt{/v2/languages} which is a simple \texttt{GET} with no parameters (and so it is trivially covered by just calling it once),
and \texttt{/v2/check} which is a \texttt{POST} with 11 input parameters.
However, looking at the source code of \texttt{ApiV2\-.handle\-Request} it seems there are some more endpoints, although those are not specified in the OpenAPI/Swagger schema.
Without an extensive analysis of this SUT, it is unclear how much of its code would not be coverable due to this issue.

\begin{figure}
\begin{lstlisting}[language=java,basicstyle=\footnotesize]
private void handleCheckRequest(
        HttpExchange httpExchange, 
        Map<String, String> parameters, 
        ErrorRequestLimiter errorRequestLimiter, 
        String remoteAddress) throws Exception {
 AnnotatedText aText;
 if(parameters.containsKey("text") && 
            parameters.containsKey("data")) {
    throw new IllegalArgumentException(
      "Set only 'text' or 'data' parameter, not both");
 } else if (parameters.containsKey("text")) {
    aText = new AnnotatedTextBuilder()
        .addText(parameters.get("text")).build();
 } else if (parameters.containsKey("data")) {
    ObjectMapper mapper = new ObjectMapper();
    JsonNode data = mapper.readTree(
                             parameters.get("data"));
\end{lstlisting}
\caption{\label{fig:language}
A function snippet from the \texttt{ApiV2} class in the API \languagetool.
}
\end{figure}

Following the execution from the entry point \texttt{/v2/check}, there are few missed branches that are worthy to discuss.
In Figure~\ref{fig:language}, the last statement crashes due to \texttt{mapper.readTree} throwing an exception (and so all statements after it cannot be executed). 
Here, when a form in a \texttt{application/x-www-form-urlencoded} payload is received, one out of the 11 parameters is treated as JSON data (i.e., the input called \texttt{data}).
Such info is written in the schema as a comment, although the type is registered as a simple string.
It is unlikely that a random string would represent a valid JSON object.

\sloppy
Another issue is when entering in the class \texttt{TextChecker}.
Several \texttt{if} statements refer to input parameters that are undefined in the schema, such as 
\texttt{textSessionId},
\texttt{noopLanguages},
\texttt{preferred\-Languages},
\texttt{enableTempOffRules},
\texttt{allow\-Incomplete\-Results},
\texttt{enable\-Hidden\-Rules},
\texttt{ruleValues},
\texttt{sourceText},
\texttt{sourceLanguage} 
and
\texttt{multi\-lingual}.
Dealing with underdefined schemas is a major issue with fuzzers, especially black-box ones.
Technically, these can be considered as \emph{``faults''} in the schemas, which should be fixed by the users.
However, schemas can be considered as \emph{documentation}, and issues in the documentations might receive less priority compared to faults in the API implementation.
Still, as fuzzers heavily rely on such schemas, a more widespread use of fuzzers might lead in changes in industrial practice by giving compelling reasons to timely update and fix those schemas.

As there are tens of thousands of line that were not covered in this SUT, there are many other missed branches that would be interesting to discuss in details. 
But it is unclear how related or independent they are from the aforementioned issues.
Once the issue of dealing of underdefined schemas is solved, it will be important to re-run these experiments to identify which major issues are still left.

\subsubsection{ocvn-rest}
On this API, with black-box testing line coverage was up to 35.3\%, which is significantly higher than the 24.8\% achieved by white-box testing.
It is the second largest API in our study, following \languagetool, based on number of lines of code.
But, in terms of endpoints, it is the largest with 192 of them.
It would not be possible, given the space, to discuss each single of these 192 endpoints. 
To analyze the achieved coverage, we hence used the test suites generated with both white-box and black-box testing.
These have hundreds of test cases.
Even calling each endpoint just once would result in at least 192 HTTP calls in the generated tests.
Considering the complexity of this SUT, maybe using a larger search budget over 1 hour could be recommended.

More than 35\% of the codebase (packages \texttt{org.devgateway.ocvn} and \texttt{org.devgateway.ocds.persistence}) deal with connections to databases such as SQL and MongoDB, but such code does not seem to be executed.
This might happen if the endpoint executions fail before writing/reading from the databases (e.g., due to input validation). 
The definitions of HTTP handlers in the package \texttt{org.devgateway.ocds.web.rest.controller}
takes more than 30\% of the codebase. 
But only half of it (i.e., around 15\%) gets covered by the tests.
Few endpoints (e.g., in the class \texttt{User\-Dashboard\-Rest\-Controller}) were not covered because they require admin authentication, which was not setup for these experiments (recall the discussion about authentication in Section~\ref{sub:bb}).
The class \texttt{Corruption\-Risk\-Dashboard\-Indicators\-Stats\-Controller} defines several HTTP endpoints, but none of them appears in the OpenAPI/Swagger schema. 
So, they cannot be called by the fuzzers.

Around 5\% of the codebase seems to deal with the generation of Excel charts, but that code is not covered.
Out of the 14 HTTP endpoints dealing with this functionality, they all fail on the very first line (and so a substantial part of the codebase is not executed, possibly much higher than 5\%), which is a variation of the following statement with different string inputs as second parameter:

\begin{lstlisting}[language=java,basicstyle=\footnotesize]
final String chartTitle = translationService.getValue(
                filter.getLanguage(),
                "charts:cancelledFunding:title");
\end{lstlisting}

Here, the language is given as input in the HTTP requests, which then gets matched with a regular expression. 
This makes the \texttt{translationService.getValue} call crash when the regular expression is not satisfied.  
White-box testing can analyze these cases, but it can be hard for black-box techniques if the info on such regular expressions is  not available in the OpenAPI/Swagger schemas.

Another large part of the code that is not covered (around another 5\%) is in the package 
\texttt{org.devgateway.ocds.web.spring}. 
However, most of this code seems only be called by the frontend of the OCVN application (and the jar file for \ocvnrest does not include such module).
From the point of view of the HTTP endpoints, this can be considered dead-code, because it cannot be reached from those endpoints. 

\begin{figure}
\begin{lstlisting}[language=java,basicstyle=\footnotesize]
@ApiOperation(value = "Count the tenders and group the results by year. The year is calculated from "
    + "tender.tenderPeriod.startDate.")
@RequestMapping(value = "/api/countTendersByYear", method = { RequestMethod.POST, RequestMethod.GET },
    produces = "application/json")
public List<DBObject> countTendersByYear(@ModelAttribute @Valid final YearFilterPagingRequest filter) {
\end{lstlisting}
\caption{\label{fig:ocvn:function}
A function snippet from the \texttt{CountPlansTendersAwardsController} class in the API \emph{ocvn-rest}.
}
\end{figure}

\begin{figure}
\begin{lstlisting}[language=java,basicstyle=\footnotesize]
@EachRange(min = MIN_REQ_YEAR, max = MAX_REQ_YEAR)
protected TreeSet<Integer> year;

@EachRange(min = MIN_MONTH, max = MAX_MONTH)
protected TreeSet<Integer> month;

@EachPattern(regexp = "^[a-zA-Z0-9]*$")
private TreeSet<String> bidTypeId;        

@EachPattern(regexp = "^[a-zA-Z0-9]*$")
private TreeSet<String> notBidTypeId;

@EachPattern(regexp = "^[a-zA-Z0-9]*$")
private TreeSet<String> procuringEntityId;

@EachPattern(regexp = "^[a-zA-Z0-9]*$")
private TreeSet<String> notProcuringEntityId;

@EachPattern(regexp = "^[a-zA-Z0-9]*$")
private TreeSet<String> contrMethod;

@Min(0)
protected Integer pageNumber;

@Range(min = 1, max = MAX_PAGE_SIZE)
protected Integer pageSize;
\end{lstlisting}
\caption{\label{fig:ocvn:constraints}
All fields with declared constraints (9 in total) in the class \texttt{YearFilterPagingRequest} in the API \emph{ocvn-rest}.
}
\end{figure}

It might be surprising that white-box testing gives significantly worse results than grey-box testing, by a large margin (i.e., 10.5\%). 
In a combinatorial optimization problem this can happen when the fitness function provides little to no gradient to the search, generating the so called \emph{fitness plateaus}.
Mutation operators that only do small changes to the chromosome of an individual (i.e., an evolved test case in this context) would have lower chances to escape from such local optima.
\evo has some advance mechanism to address these cases, like for example \emph{adaptive hypermutation}~\cite{zhang2021adaptive}.
However, it was not enough to handle \emph{ocvn-rest}.
The problem here is that most endpoints take a JSON object as input, with several constraints that must be satisfied. 
But those constraints are not evaluated in the business logic of the SUT, but rather in a third-party library. 
These constraints  are not specified in the OpenAPI schema.
Figure~\ref{fig:ocvn:function} shows an example, in which the JSON input is unmarshalled into an \texttt{YearFilterPagingRequest} instance.
Because this input is marked with the annotation \texttt{@Valid}, its validity is evaluated before the method \texttt{countTendersByYear()} is called. 
If it is not valid, then the Spring framework returns an HTTP response with status 400 (i.e., user error), without calling the method \texttt{countTendersByYear()}.

As shown in Figure~\ref{fig:ocvn:constraints}, the class \texttt{YearFilterPagingRequest} has 9 fields on which \texttt{javax.validation} constraints are declared. 
Only need 1 (out of 9) violated constraint to invalidate the whole object. 
Because the constraints are evaluated in a third-party library, the fitness function has no gradient to guide the search to solve all of them.
Looking at the results, it seems like that generating valid instances at random is feasible, albeit with low probability.
On the other hand, small mutations on an existing invalid instance have little to no chance to make the object valid. 

To address these issues, there could be at least three strategies:

\begin{itemize}
\item (White-Box) Compute the fitness function to optimize the coverage (using different search heuristics) on all the code, including third-party libraries, and not just the business logic of the SUT.
In this way, there would be gradient for generating tests in which the code that evaluates \texttt{@Valid} constraints returns true. 
Although in theory possible, this does not sound like a very promising approach (unless only the needed code in the third-party libraries is instrumented).
In most cases, the code of the business logic of the SUT is just tiny compared to the source code of all used third-party libraries, which includes application frameworks such as Spring, HTTP servers such as Tomcat, and ORM libraries such as Hibernate. 
Instrumenting everything to compute advance heuristics on all third-party libraries would likely have huge performance overhead.
This is particularly the case for all code that is executed before the business logic, e.g., the unmarshalling of incoming HTTP requests. 

\item (White-Box) Natively support \texttt{@Valid} in the fitness function, by creating a branch distance for each field, with the objective of minimizing the distance for all fields. 
Different branch distances can be computed for the different constraints which are part of the standard \texttt{javax.validation} (e.g., \texttt{javax.validation.constraints.Min}).
A new testing target (to optimize for) could be created for each \texttt{@Valid} annotation in the business logic of the SUT. 
However, one major challenge here is that \texttt{javax.validation} allows to create custom constraint annotations.
This is for example the case for \texttt{@EachPattern} and \texttt{@EachRange} used in Figure~\ref{fig:ocvn:constraints}.

\item (Black-Box) Even if the info on the field constraints are missing in the OpenAPI schema, it can be detected in the schema that the same object type is used as input in more than one endpoint.
If that is the case, and if for an endpoint it is difficult to create a valid instance (e.g., all evaluated HTTP calls so far return a response with HTTP status 400), then a possible strategy could be to re-use as input an instance created for another endpoint for which a 2xx status code was returned (if any has been created so far during the search). 

\end{itemize}

\subsubsection{realworld-app}

\begin{figure}
\begin{lstlisting}[language=java,basicstyle=\footnotesize]
unFavorite(id, slug) {
   return __awaiter(this, void 0, void 0, function* () {
      let article = yield this.articleRepository.findOne({ slug });
      const user = yield this.userRepository.findOne(id);
      const deleteIndex = user.favorites.findIndex(_article => _article.id === article.id);
      if (deleteIndex >= 0) {
\end{lstlisting}
\caption{\label{fig:real}
A function snippet from the \texttt{article.service.js} file in the API \emph{realworld-app}.
}
\end{figure}

On this API, there is a small difference between \evo black-box and \Schemathesis (c8 instrumentation: 69.4\% vs.~69.7\%),
and between grey-box and white-box \evo (\evo instrumentation without bootstrap coverage: 24.6\% vs.24.4\%).
 
Most of the code is covered.
What is left are either unfeasible branches, or branches related to fetching data from the database (MySQL in this case) with some given properties. 
Figure~\ref{fig:real} shows one such example, where 3 queries into the database are executed, based on the 2 inputs \texttt{id} and \texttt{slug}.
The condition of \texttt{if} statement at the end of that code snippet is never satisfied, as it depends on these 3 SQL queries.   
Recall that, although \evo has support for SQL query analyses (which could help in this case), it is currently only for the JVM, and not NodeJS.

\subsubsection{proxyprint}
With a line coverage of up to 53.3\%, this is an API in which reasonable results are achieved, but more could be done.
Several functions in this API are never called, and so they are impossible to cover,
like for example the methods \texttt{send\-Email\-Finished\-Print\-Request} 
and \texttt{sendEmailCancelledPrintRequest} in the class \texttt{Mail\-Box},
and the private methods \texttt{singleFileHandle} and \texttt{calc\-Budgets\-For\-Print\-Shops}
in the class \texttt{ConsumerController}.
So, 100\% coverage is not possible on this API.

\begin{figure}
\begin{lstlisting}[language=java,basicstyle=\footnotesize]
@Secured({"ROLE_USER"})
@RequestMapping(value = "/consumer/requests", 
                method = RequestMethod.GET)
public String getRequests(Principal principal) {
   JsonObject response = new JsonObject();
   Consumer consumer = consumers.findByUsername(
                             principal.getName());
   if(consumer == null) {
        response.addProperty("success", false);
        return GSON.toJson(response);
   }
\end{lstlisting}
\caption{\label{fig:proxyprint:auth}
Snippet of function handler for the endpoint \texttt{/consumer/request} in \proxyprint.
}
\end{figure}

This API provides a selection of various branches that are not covered.
However, it is not straightforward to find out the interesting challenges here.
One problem is that the number of HTTP calls on this API is low, and so not much search is actually carried out compared to the other SUTs when using a 1 hour budget.
For example,  the generated statistic files of \evo report an average of 37 393 HTTP calls per experiment on \proxyprint, whereas for example for \restnews it is 372 733, i.e., nearly 10 times as more.
So, many of these missed branches could be covered if running \evo for longer.
Still, some of these branches seem quite unlikely to be covered even with larger search budgets.
Let us discuss a few of them.
Some are related to authentication.
For example, in Figure~\ref{fig:proxyprint:auth}, the code inside the \texttt{if} statement is never executed.
To reach that statement, an HTTP call with valid authentication information needs to be provided. 
Such authentication info is validated with the database when the \texttt{Principal} object is instantiated by the framework (Spring Security in this case). 
Authentication information needs to be available when the SUT starts (unless the API has some way to register new users directly on-the-fly from the API itself).
Therefore, to deal with security (especially when involving hashed passwords), some initialization script is required, e.g., to register a set of users with valid username/password information.
In the case of \proxyprint, also some other data in other tables is created as well, e.g., in the \texttt{Consumer} table.
To be able to cover such branch, either a fuzzer should find a way to create new valid users, or modify the existing data in the database (e.g., \evo can add new data, but not modify the existing one~\cite{arcuri2020handling}).

\begin{figure}
\begin{lstlisting}[language=java,basicstyle=\footnotesize]
@RequestMapping(value="paypal/ipn/consumer/{consumerID}",
                method=RequestMethod.POST)
protected void consumerLoadUpConfirmation(
      @PathVariable(value = "consumerID") long cid, 
      HttpServletRequest request, 
      HttpServletResponse response
      ) throws ServletException, IOException {
  Map<String,String> configurationMap = 
                           Configuration.getConfig();
  IPNMessage ipnlistener = new IPNMessage(
                                     request,
                                     configurationMap);
  boolean isIpnVerified = ipnlistener.validate();
  String transactionType = ipnlistener
                                 .getTransactionType();
  Map<String,String> map = ipnlistener.getIpnMap();

  String payerEmail = map.get("payer_email");
  Double quantity = Double.valueOf(map.get("mc_gross"));
\end{lstlisting}
\caption{\label{fig:proxyprint:params}
Snippet of function handler for the endpoint \texttt{/paypal/ipn/consumer/\{consumerID\}} in \proxyprint.
}
\end{figure}

Figure~\ref{fig:proxyprint:params} shows an example in which the line defining the variable \texttt{quantity} throws an exception, due to \texttt{Double.valueOf} being called on a null input.
Here, an HTTP object \texttt{request} is passed as input to the constructor of \texttt{IPNMessage}, which is part of PayPal SDK library. 
Inside such library, \texttt{request.getParameterMap()} is called to extract all the parameters of the HTTP request, which are used to populate the map object returned by \texttt{ipnlistener.getIpnMap()}.
However, as such parameters are read dynamically at runtime, the OpenAPI/Swagger schema has no knowledge of them (as for this SUT the schema is created automatically with a library when the API starts).
Therefore, there is no info to use an HTTP parameter called \texttt{mc\_gross} of type double.
As such parameter is used directly without being transformed, testability transformations with taint analysis might be able to address this problem~\cite{arcuri2021enhancing}, but such techniques would need to be extended to support \texttt{getParameterMap()} and \texttt{Map.get}.

\begin{figure}
\begin{lstlisting}[language=java,basicstyle=\footnotesize]
@Secured("ROLE_MANAGER")
@RequestMapping(value="/printshops/{id}/pricetable/rings", method=RequestMethod.PUT)
public String editRingsItem(
         @PathVariable(value = "id") long id, 
         @RequestBody RingTableItem rti) {
    PrintShop pshop = printshops.findOne(id);
    JsonObject response = new JsonObject();

    if(pshop!=null) {
      BindingItem newBi = new BindingItem(
                   Item.RingType.valueOf(
                               rti.getRingType()), 
                               rti.getInfLim(), 
                               rti.getSupLim());
\end{lstlisting}
\caption{\label{fig:proxyprint:enum}
Snippet of function handler for the endpoint \texttt{/printshops/{id}/pricetable/rings} in \proxyprint.
}
\end{figure}

Figure~\ref{fig:proxyprint:enum} shows another interesting example, where the value of a string field in a JSON payload is used directly to instantiate an enum value, i.e., the statement \texttt{Item.RingType.\-valueOf(\-rti\-.get\-Ring\-Type())}.
This fails, as a random string is extremely unlikely to represent a valid value from a restricted set. 
This might be handled by providing such info in the OpenAPI/Swagger schema (which supports defining enumerations on string fields), or also by handling \texttt{valueOf()} in enumeration  by extending the techniques in~\cite{arcuri2021enhancing} to support it.

The handler \texttt{calc\-Budget\-For\-Print\-Request} for the endpoint \texttt{/consumer\-/budget} is never called (and so all the business logic related to this endpoint remains uncovered by the tests).
Such endpoint requires as input a payload with type \texttt{multipart/form-data}, but the schema wrongly specifies the \texttt{application/json} type.
A fuzzer might be able to handle this case automatically, but it could be considered as a major issue in the schema, that should be fixed by the users like a fault in the SUT.
In other words, there is a big difference between not providing all information (e.g., missing enum value constraints like in the case of \texttt{/printshops\-/\{id\}\-/pricetable\-/rings}) and providing \emph{wrong} information (i.e., wrong payload type as in the case of \texttt{/consumer/budget}).

\subsubsection{rest-ncs}
\label{sub:sut:ncs}

On this SUT, white-box testing with \evo achieves an average of 93\%.
An analysis of the non-covered lines shows that those are not possible to reach, i.e., dead-code.
An example is checking twice if a variable is lower than 0, and return an error if so.
In such case, it is impossible to make the second check true. 
Therefore, as the maximum achievable coverage is obtained in each single run, this can be considered as a \emph{solved} problem.  

Regarding black-box testing, all but 2 fuzzers achieve more than 50\% line coverage, with \Schemathesis achieving 94.1\% coverage (computed with JaCoCo), followed by \RestCT achieving 85.5\% average coverage.
On this API, black-box \evo gives worse results, i.e., 64.4\%.

This is a rather interesting phenomenon, which is strongly dependent on some design choices these fuzzers make.
For example, some branches in this API depend on whether 2 or more integer inputs are equal (e.g., in the case of Triangle Classification).
Given two 32-bit integers $X$ and $Y$, there is only a $1/2^{32}$ chance that $X=Y$.
On average, it will need to sample more than 2 billion values before obtaining $X=Y$.
With such constraints, it will be very difficult for a black-box fuzzer to cover this kind of branches if integer inputs are sampled at random.
However, a fuzzer does not need to sample from the whole spectrum of all possible integers.
For example, it could sample from a restricted range, e.g., $[-100,100]$.
On the one hand, the shorter the range, the highest chances to sample $X=Y$.
On the other hand, a short range could make impossible to cover branches that require values outside such range (e.g., 12345). 
This is a tradeoff that needs to be made.

\subsubsection{rest-news}
On this API, it was possible to achieve up to 66.5\% line coverage.  
As for the other SUTs, most of the missing coverage is due to dead-code, which is impossible to execute. 
However, there are some branches that should be possible to cover, but they were not. 
These seem all related to the same issue, which is the update of existing data in the database.
Databases allow to define constraints on their data (e.g., a number should be within a certain range), using SQL commands such as \texttt{CHECK}.
White-box \evo can generated test data directly into SQL databases~\cite{arcuri2020handling}, taking into account all these constraints (as all this info is available in the schema of the database).
If any constraint is not satisfied, then the \texttt{INSERT} SQL commands will fail.
The problem here is that the SUT might define further constraints on such data.
In JVM projects, this is commonly done with \texttt{javax.validation} annotations when using ORM libraries such as Hibernate~\cite{Hibernate}.
But \evo does not seem to handle this, and generates data that is valid for the database (as all the \texttt{CHECK} operations pass and the \texttt{INSERT}s do not fail), but not for these further constraints.
So, in an update endpoint, invalid (from the point of view of \texttt{javax.validation} constraints) data can be read from the database, which then the SUT fails to write it back (as these \texttt{javax.validation} constraints are check on each write operation) when the update endpoint has done its computations.

\begin{figure}
\begin{lstlisting}[language=java,basicstyle=\footnotesize]
public static String  subject(String directory, String file){
		int result = 0;
		String[] fileparts = null;
		int lastpart  = 0;
		String suffix = null;  
		fileparts = file.split(".");
		lastpart = fileparts.length - 1;
		if (lastpart > 0) {
\end{lstlisting}
\caption{\label{fig:scs-rest}
Snippet of \texttt{subject} function from the \texttt{FileSuffix} class in the API \emph{rest-scs}.
}
\end{figure}  
 
\subsubsection{rest-scs}
\label{sub:sut:scs}

With a line coverage of up to 86.2\%, this is an API in which white-box testing is highly effective. 
There were two types of branches that were not covered: the ones involving regex checks with \texttt{Pattern.matches}, and the other involving the dot character ``.''. 
In this latter case, it is due to \evo not using such character when sampling random strings, which make the \texttt{if} statement in Figure~\ref{fig:scs-rest} impossible to cover.
This might sound like something rather simple to fix.
However, as soon as having special characters used in random strings, extra care needs to be taken into consideration when outputting test cases (e.g., the character `\$' has special meaning in Kotlin, and would need to be escaped when used in strings, which is not the case for Java).   
Character escaping rules can be very different based on the context in which they are used in a test case (e.g., inside a URL, inside a JSON object passed as HTTP body payload or data injected into a SQL database).
All these cases have to be handled, otherwise the tools would end up generating test cases that do not compile.

\subsubsection{restcountries}

\begin{figure}
\begin{lstlisting}[language=java,basicstyle=\footnotesize]
List<Country> countries = CountryService.getInstance().getByCodeList(codes);
if (!countries.isEmpty()) {
    return parsedCountries(countries, fields);
}
return getResponse(Response.Status.NOT_FOUND);
\end{lstlisting}            
\caption{\label{fig:restcountries}
Code snippet from the \texttt{CountryRestV2} class in the API \restcountries.
}
\end{figure}  

On this API, high coverage is achieved (i.e., 77.1\%).
Like the other SUTs, there is quite a bit of dead-code due to getters/setters that are never called and catch blocks for exceptions that might not be possible to throw via test cases.
The class \texttt{StripeRest} presents an interesting case, as it defines the endpoint \texttt{/contribute}, but its info is not in the schema (and so such endpoint is never called).

Figure~\ref{fig:restcountries} shows a code snippet which is repeated several times with small variations in few endpoints.
Here, this API returns a list of countries based on different filtering criteria, like for example the country codes.
However, the response with \texttt{NOT\_FOUND} is never returned.
The problem is that, even if the HTTP requests provide invalid inputs (e.g., a country code that does not exist), the \texttt{countries} list is wrongly populated with null values, and so the list is never empty.
This is an interesting bug in the API, which then  result in a crash (i.e., a returned 500 status code), as \texttt{parsedCountries} throws an exception.
However, until this fault in the API is fixed, all these statements with \texttt{NOT\_FOUND} are technically dead-code that cannot be reached with the tests.

Similarly to \restncs, the fuzzing of this API might be considered as a \emph{solved} problem, at least for its current faulty version.

\begin{figure}
\begin{lstlisting}[language=java,basicstyle=\footnotesize]
@POST @Timed @UnitOfWork
@Consumes(MediaType.APPLICATION_JSON)
public MediaFile create(
          @Auth @ApiParam(hidden = true) AuthResult authResult, 
          @Context HttpServletResponse response,
          MediaFile mediaFile) {
   doAuth(authResult, response, Permission.mediaitem_create);
   try {
   URI uri = new URI(mediaFile.getUri());
   if ("data".equals(uri.getScheme())) {
\end{lstlisting}
\caption{\label{fig:scout-api}
Snippet of \texttt{create} function from the \texttt{MediaFileResource} class in the API \emph{scout-api}.
}
\end{figure}

\subsubsection{scout-api}
On this API, white-box \evo achieves slightly worse results (53.4\% vs.~54.7\%), although the difference is not statistically different (based on 10 runs).  
Large part of its codebase (more than a third) cannot be executed.
For example, \emph{scout-api} can start in the background a thread to fetch and analyse some data, independently from the RESTful endpoints (i.e., the whole module \texttt{data-batch-jobs}).
Such thread is deactivated by default, and so nearly 30\% of the whole codebase is not executed.
Another non-negligible part of the codebase (around 5\%) is related to authentication using OAuth via Google APIs (\emph{scout-api} provides different ways to authenticate). 

There is a large part of the codebase that could be executed, but it is not due to an \texttt{if} statement at the beginning of one of the HTTP endpoints.
This is shown in Figure~\ref{fig:scout-api}.
Here an incoming JSON payload is unmarshalled into a \texttt{MediaFile} object, which has a string field that is treated as a URI.
Most random strings are a valid URI, as a URI can just be a name.
But, it is extremely unlikely that a random string would represent a URI with a valid scheme component, and so \texttt{uri.getScheme()} returns null. 
However, this is a case that likely would be trivial to solve with taint analysis, e.g., by extending the techniques in~\cite{arcuri2021enhancing} to support URI objects.
This does not mean though that much higher coverage would be achieved, it depends on all the remaining code that is executed once that predicate is satisfied. 

Even without code analyses, this type of challenge could also be addressed with black-box techniques. 
For example, in this API the string field representing the URI is called \texttt{uri}.
An analysis of the names of the fields can be used to possibly infer their expected type.
For example, any field whose name starts or ends with \texttt{uri} could be treated as an actual URI when data is generated. 
Also, some types are quite common in RESTful APIs, like strings representing URLs and dates.
Instead of sampling strings completely at random, it could make sense to sample with some probability strings with given types that are commonly used.
This might now work in all cases, but it is something worth to investigate.

\subsubsection{spacex-api}

Similarly to \emph{realworld-app}, in this API there is only a small difference between \evo black-box and \Schemathesis (c8 instrumentation: 84.7\% vs.~85.4\%),
and between grey-box and white-box \evo (\evo instrumentation without bootstrap coverage: 41.5\% vs.~41.5\%, but with $\hat{A}_{12}=0.58$).
There are two main issues affecting the coverage results here. 

First, this API requires authentication, where each API endpoint requires specific authorization roles to be able to be executed.
For these experiments, a user with the right credentials was created in the database, and the fuzzers were given the authorization info to send HTTP messages on behalf of this user.
The setting up of this user was done manually, in a script.
However, after running the experiments and analyzing the generated tests, we realized that some authorization roles were missing (e.g., \texttt{capsule:create}). 
So, few endpoints were not covered due to misconfigured authentication.

\begin{figure}
\begin{lstlisting}[language=java,basicstyle=\footnotesize]
router.get('/:id', cache(300), async (ctx) => {
  const result = await Core.findById(ctx.params.id);
  if (!result) {
    ctx.throw(404);
  }
  ctx.status = 200;
  ctx.body = result;
});
\end{lstlisting}
\caption{\label{fig:spacex:get}
Snippet of the handle for the \texttt{GET /cores/:id} endpoint in the API \emph{spacex-api}.
}
\end{figure}

\begin{figure}
\begin{lstlisting}[language=java,basicstyle=\footnotesize]
router.post('/', auth, authz('core:create'), async (ctx) => {
  try {
    const core = new Core(ctx.request.body);
    await core.save();
    ctx.status = 201;
  } catch (error) {
    ctx.throw(400, error.message);
  }
});
\end{lstlisting}
\caption{\label{fig:spacex:post}
Snippet of the handle for the \texttt{POST /cores} endpoint in the API \emph{spacex-api}.
}
\end{figure} 

The second  main issue is related to the database. 
Figure~\ref{fig:spacex:get} shows an example in which a record is search by id. 
No test generated by \evo was able to make a call that returned a 200 HTTP status code.
For white-box testing, \evo fails to do this because it is not able to analyze queries on MongoDB databases.
Even with black-box testing, it could be possible to create a new record with a \texttt{POST}, and use its id for a following \texttt{GET}.
Fuzzers can exploit this kind of information to create sequences of HTTP calls where ids are linked.
But, this all depends on how the link is defined.
Figure~\ref{fig:spacex:post} shows the implementation of the \texttt{POST} endpoint to create such record.
\evo has no issue to fully cover the code of such endpoint, and create valid records.
Also, \evo uses different strategies to \emph{link} endpoints that manipulate the same resources~\cite{arcuri2019restful,zhang2021resource}.
But those are based on  the recorded interactions with the database (not done here for NodeJS and MongoDB), and on  best practices in RESTful API design.
For example, it is a recommended practice that \texttt{POST}s on collections of resources (e.g., \texttt{/cores}) create a new resource, whose \texttt{id} (chosen on the server, to guarantee it is unique, unless it is a UUID) is present in the \texttt{location} HTTP header of the response (e.g., \texttt{location: /cores/42}).
\evo can use these location headers to create \texttt{POST} requests followed with the appropriate linked \texttt{GET}.
However, on this API, the \texttt{POST} implementation in Figure~\ref{fig:spacex:post} does not seem to follow best practices in REST API design.
For example,  it does not setup the \texttt{location} header in the response, and neither it returns the generated \texttt{id} (e.g., in the body payload).
However, it could still be possible to test this kind of API by doing the following: create a new resource with \texttt{POST /cores}, followed by a \texttt{GET} of the whole collection (i.e., \texttt{GET /cores}), and then extract the \texttt{id} fields from this response to call \texttt{GET /cores/:id} with a valid \texttt{id}.
Note: this is assuming that the collection is empty.
If it is not, then the first \texttt{POST} is not even necessary.
However, this also assumes that the name of the field representing the id is the same (or at least very similar) in both the body payload and endpoint path parameter (so they can be matched).

%-----------------------------------------------------------
\subsection{Discussion}
\label{sub:discussion}

Based on the analyses of the logs and tests generated for the 19 APIs, some general observations can be made:

\begin{itemize}

\item Many research prototypes are not particularly robust, and can crash when applied on new SUTs.
For example, we have faced this issue with \evo many times, like when adding a new API to EMB~\cite{EMB}.
Although we add new APIs to EMB each year, EMB has been available as open-source since 2017, and anyone can use it for their empirical studies and make sure their tools do not crash on it. 
However, it is important to stress out that, in this paper, we have compared \emph{tool implementations} rather than \emph{techniques}.
For example, a tool with low performance (e.g., due to crashes) could still feature novel techniques that could be very useful, e.g., if integrated or re-implemented in a more mature tool.

\item Like software, also schemas can have faults, and/or omissions (e.g., constraints on some inputs might be missing). And this issue seems quite common, especially when schemas are written manually. 
Although this problem could be addressed by white-box testing (currently supported only by \evo) by analyzing the source code of the SUTs, it looks like a major issue for black-box testing, which might not have a viable solution.

\item Interactions with databases are common in RESTful APIs. To execute the code to fetch some data, such data should be first present in the database. The data could be created with endpoints of the API itself (e.g., \texttt{POST} requests), as well as inserted directly into the database with SQL commands.
Both approaches have challenges: e.g., how to properly link different endpoints that work on the same resources, and how to deal with data constraints that are specified in the code of the API and not in the SQL schema of the database.
The compared fuzzers provide different solutions to address this problem, but it is clear that more still need to be done.

\item Currently, no fuzzer deals with the mocking of external web services (e.g., using libraries such as WireMock). Testing with external live services has many shortcomings (e.g., the generated tests can become flaky), but it might be the only option for black-box strategies.  For white-box strategies, mocking of web services will be essential when testing industrial enterprise systems developed with a microservice architecture (as in this class of software interactions between web services are very common).

\end{itemize}

\begin{result}
{\bf RQ3:} There are several open challenges in fuzzing RESTful APIs, including for example how to deal with underspecified schemas, and how to deal with interactions with external services (e.g., databases and other APIs).  
\end{result}

%%%%%%%%%%%%%%%%%%%%%%%%%%%%%%%%%%%%%%%%%%%%%%%%%%%%%%%%%%%%%%%%%%%%%%%%%%%%
\section{Threats To Validity}
\label{sec:threats}

{\bf Internal Validity}.
Besides writing the scaffolding to run and analyze the experiments,
before running the experiments 
we did not make code modifications in any existing tool for this study.
Although we are the authors of \evo, to try to be fair, we used the latest release before running these experiments, without re-running them after fixing any new issues.
However, as we have used all these APIs in our previous studies, \evo did not crash on any of them. 

We used seven existing tools, with their latest release versions when possible.
However, there is a possibility that we might have misconfigured them, especially considering that some of those tools have minimal documentation.
To avoid such possibility, we carefully look at their execution logs, to see if there was any clear case of misconfiguration.
Furthermore, we release all our scripts as open-source in the repository of \evo, so anyone can review them, and replicate the study if needed.

Any comparison of tools made by the authors of one of these tools is bound to be potentially bias, especially if such tool turns out to give the best results (as in our case with \evo).
However, the relative performance of the other six tools among them would be not affected by this issue.
Likewise, the in-depth analysis of the SUTs is not affected by this issue either. 

All the compared fuzzers use randomized algorithms.
To take this into account, each experiment was repeated 10 times, and we analyzed them with the appropriate statistical tests.

{\bf External Validity}.
The chosen seven fuzzers are arguably representing the state-of-the-art in testing RESTful APIs. 
However, results on 19 APIs might not generalize to other APIs as well, although we selected different programming runtimes (e.g., JVM and NodeJS) as well including one industrial API. 
This kind of system testing experiments are expensive (nearly 71.2 days of computational effort in our case), which makes using more APIs challenging.
Furthermore, finding RESTful APIs in open-source repositories is not so simple, and considerable effort might then be needed to configure them (e.g., setup external dependencies like databases and find out how/if they use any form of authentication).

%%%%%%%%%%%%%%%%%%%%%%%%%%%%%%%%%%%%%%%%%%%%%%%%%%%%%%%%%%%%%%%%%%%%%%%%%%%%
\section{Conclusions}
\label{sec:conclusions}

RESTful APIs are widely used in industry, and so several techniques have been developed in the research community to automatically test them.
Several reports in the literature show the usefulness in practice of these techniques, by reporting actual faults automatically found with these fuzzers.
However, not much has been reported on how the different techniques compare, nor on how effective they are at covering different parts of the code of these APIs.
This latter is usually the case due to the testing of remote APIs, for which researchers do not have access to their source code.

To address these issues, in this paper we have compared the state-of-the-art in fuzzing RESTful APIs, using seven fuzzers to test 19 APIs, totaling more than 280 thousands of lines of code (for their business logic, and not including the millions of lines of code of all the third-party libraries they use). 
Each fuzzer was run for 1 hour, and each experiment was repeated 10 times, to take into account the randomness of these tools. 

The results show different degrees of coverage for the different black-box testing tools, where \evo seems the tool giving the best results (i.e., highest code coverage on average) on this case study, closely followed by \Schemathesis.
However, no black-box fuzzer was able to achieve more than 60\% line coverage on average.
Furthermore, the experiments show that white-box testing gives better results  than black-box testing.  
Still, large parts of these APIs are left uncovered, as the fuzzers  do not manage to generate the right data to maximize code coverage. 
Although these fuzzers are already useful for practitioners in industry, more needs to be done to achieve better results.

To address this issue,
this large empirical analysis has then be followed by an in-depth analysis of the source code of each of these 19 APIs, to understand what are the main issues that prevent the fuzzers from achieving better results. 
Several issues were identified, including 
for example how to deal with underspecified schemas, and how to deal with interactions with external services (e.g., other APIs and databases).
This provides a useful list of common problems that researchers can use to drive new research effort in improving performance on this problem domain.

For few of these issues, we discussed possible solutions, but those will need to be empirically validated. 
Future work will aim at using the new knowledge and insight provided in this paper to design new variants of these tools, to be able to achieve better results.
In particular, the insight provided in this study is shaping the current research efforts in \evo, pointing to clear research issues that need to be prioritized.

All the tools and APIs used in this study are available online.
To enable the replicability of this study, all our scripts used in our experiments are published as open-source, available online in the repository of \evo, at \url{www.evomaster.org}.
These scripts get automatically stored on Zenodo at each new release (e.g., version $1.5.0$~\cite{andrea_arcuri_2022_6651631}).

%----------------------------------------------------------------------------------------------------------
\section*{Acknowledgments}
This work is funded by the European Research Council (ERC) under the European Union’s Horizon 2020 research and innovation programme (EAST project, grant agreement No. 864972).

%https://arxiv.org/help/submit_tex#latex
% USE generated bbl

%\bibliographystyle{ACM-Reference-Format}
%\bibliography{../../../papers}
\input{rest-comparisons.bbl}

\end{document}

%% file: generated_files/sut_info.tex
\begin{tabular}{  l l r r r }\\ 
\toprule 
SUT & Language & Files & File LOCs & c8/JaCoCo LOCs \\ 
\midrule 
\emph{cyclotron}  & JavaScript & 25 & 5803 & 2458\\ 
\emph{disease-sh-api}  & JavaScript & 57 & 3343 & 2997\\ 
\emph{js-rest-ncs}  & JavaScript & 8 & 775 & 768\\ 
\emph{js-rest-scs}  & JavaScript & 13 & 1046 & 1044\\ 
\emph{realworld-app}  & TypeScript & 37 & 1229 & 1077\\ 
\emph{spacex-api}  & JavaScript & 63 & 4966 & 3144\\ 
\hline 
\emph{catwatch}  & Java & 106 & 9636 & 1835\\ 
\emph{cwa-verification}  & Java & 47 & 3955 & 711\\ 
\emph{features-service}  & Java & 39 & 2275 & 457\\ 
\emph{gestaohospital-rest}  & Java & 33 & 3506 & 1056\\ 
\emph{ind0}  & Java & 75 & 5687 & 1674\\ 
\emph{languagetool}  & Java & 1385 & 174781 & 45445\\ 
\emph{ocvn-rest}  & Java & 526 & 45521 & 6868\\ 
\emph{proxyprint}  & Java & 73 & 8338 & 2958\\ 
\emph{rest-ncs}  & Java & 9 & 605 & 275\\ 
\emph{rest-news}  & Kotlin & 11 & 857 & 144\\ 
\emph{rest-scs}  & Java & 13 & 862 & 295\\ 
\emph{restcountries}  & Java & 24 & 1977 & 543\\ 
\emph{scout-api}  & Java & 93 & 9736 & 2673\\ 
\midrule 
\emph{Total} & & 2637(203,2434) & 284898(17162,267736) & 76422(11488,64934)\\ 
\bottomrule 
\end{tabular} 

%% file: generated_files/tableBB_1h.tex
\begin{tabular}{ l  r r r r r r r    }\\ 
\toprule 
SUT  &  \bBOXRT &  \evo BB &  \RestCT &  \Restler &  \RestTest &  \RestTestGenVT &  \Schemathesis \\ 
\midrule 
\emph{cyclotron} & 41.3 [41.3,41.3] (5) & {\bf 70.3 [70.1,72.3] (1)} & 41.3 [41.3,41.3] (5) & 41.3 [41.3,41.3] (5) & 41.3 [41.3,41.3] (5) & 41.3 [41.3,41.3] (5) & 69.1 [67.7,69.7] (2) \\ 
\emph{disease-sh-api} & 56.4 [55.4,57.4] (3) & 60.8 [60.8,60.9] (2) & 48.4 [48.4,48.4] (6.5) & 48.5 [48.5,48.5] (4) & 48.5 [48.4,48.5] (5) & 48.4 [48.4,48.4] (6.5) & {\bf 61.5 [61.4,61.6] (1)} \\ 
\emph{js-rest-ncs} & 70.2 [67.3,71.1] (5) & 93.0 [89.8,95.8] (2) & 92.2 [87.5,92.7] (3) & 44.3 [44.3,44.3] (6.5) & 44.3 [44.3,44.3] (6.5) & 88.6 [85.8,93.0] (4) & {\bf 100.0 [100.0,100.0] (1)} \\ 
\emph{js-rest-scs} & 83.2 [83.0,83.5] (5) & {\bf 88.7 [87.5,89.5] (1)} & 83.2 [83.1,83.2] (6) & 54.1 [54.1,54.1] (7) & 84.1 [83.3,84.6] (4) & 86.4 [85.4,87.1] (2) & 86.1 [85.9,87.4] (3) \\ 
\emph{realworld-app} & 64.2 [62.8,66.4] (4) & 69.4 [69.4,69.4] (2) & 59.7 [59.7,59.7] (6) & 66.5 [66.5,66.5] (3) & 59.7 [59.7,59.7] (6) & 59.7 [59.7,59.7] (6) & {\bf 69.7 [69.1,69.8] (1)} \\ 
\emph{spacex-api} & 76.1 [76.1,76.2] (6) & 84.7 [84.7,84.8] (2) & 76.1 [76.1,76.2] (7) & 76.3 [76.3,76.4] (3) & 76.3 [76.3,76.3] (5) & 76.3 [76.3,76.4] (4) & {\bf 85.4 [85.3,85.6] (1)} \\ 
\hline 
\emph{catwatch} & 31.0 [29.2,31.8] (3) & {\bf 35.9 [34.3,36.9] (1)} & 9.7 [9.7,9.7] (7) & 17.3 [14.5,20.5] (5) & 20.8 [15.9,23.8] (4) & 15.1 [12.3,18.3] (6) & 35.8 [33.6,39.1] (2) \\ 
\emph{cwa-verification} & 43.4 [42.9,43.5] (4) & 49.4 [49.1,49.6] (2) & 21.9 [21.9,21.9] (6) & 21.9 [21.9,21.9] (6) & 21.9 [21.9,21.9] (6) & 43.8 [43.3,43.9] (3) & {\bf 49.5 [49.5,49.5] (1)} \\ 
\emph{features-service} & 35.7 [35.7,35.7] (4) & {\bf 59.7 [58.4,62.1] (1)} & 21.0 [21.0,21.0] (6) & 21.0 [21.0,21.0] (6) & 21.0 [21.0,21.0] (6) & 45.9 [45.1,46.8] (3) & 52.0 [46.6,57.1] (2) \\ 
\emph{gestaohospital-rest} & 36.2 [36.2,36.2] (4) & 50.8 [44.8,54.3] (3) & 19.9 [19.9,19.9] (7) & 21.5 [21.5,21.5] (6) & 29.0 [28.1,32.0] (5) & 57.2 [51.7,58.7] (2) & {\bf 58.7 [55.9,62.3] (1)} \\ 
\emph{ind0} & {\bf 8.2 [8.2,8.2] (3)} & {\bf 8.2 [8.2,8.2] (3)} & 7.6 [7.6,7.6] (6.5) & 7.6 [7.6,7.6] (6.5) & {\bf 8.2 [8.2,8.2] (3)} & {\bf 8.2 [8.2,8.2] (3)} & {\bf 8.2 [8.2,8.2] (3)} \\ 
\emph{languagetool} & 1.7 [1.7,1.7] (6) & {\bf 32.6 [26.0,35.1] (1)} & 1.5 [1.5,1.5] (7) & 1.9 [1.9,1.9] (4.5) & 2.5 [2.5,2.5] (2) & 1.9 [1.9,1.9] (4.5) & 2.2 [2.1,2.5] (3) \\ 
\emph{ocvn-rest} & 10.1 [10.1,10.1] (5) & {\bf 27.5 [27.5,27.6] (1)} & 10.1 [10.1,10.1] (5) & 10.1 [10.1,10.1] (5) & 10.1 [10.1,10.1] (5) & 10.1 [10.1,10.1] (5) & 27.5 [27.5,27.7] (2) \\ 
\emph{proxyprint} & 4.2 [4.2,4.2] (5) & {\bf 34.0 [32.5,35.0] (1)} & 4.2 [4.2,4.2] (5) & 4.2 [4.2,4.2] (5) & 4.2 [4.2,4.2] (5) & 4.2 [4.2,4.2] (5) & 4.4 [4.4,4.4] (2) \\ 
\emph{rest-ncs} & 55.0 [52.4,56.4] (5) & 64.5 [64.4,64.7] (3) & 85.5 [85.5,85.5] (2) & 40.7 [40.7,40.7] (6) & 5.1 [5.1,5.1] (7) & 64.3 [64.0,64.7] (4) & {\bf 94.1 [93.1,94.5] (1)} \\ 
\emph{rest-news} & 34.9 [34.0,36.8] (5) & {\bf 69.4 [69.4,69.4] (1)} & 13.9 [13.9,13.9] (6.5) & 44.4 [44.4,44.4] (4) & 13.9 [13.9,13.9] (6.5) & 47.2 [47.2,47.2] (3) & 68.8 [67.4,70.8] (2) \\ 
\emph{rest-scs} & 60.2 [59.3,61.4] (6) & {\bf 66.9 [64.4,70.2] (1)} & 60.5 [60.3,61.0] (5) & 58.3 [58.3,58.3] (7) & 61.7 [61.4,62.4] (4) & 65.3 [64.4,67.1] (2) & 64.8 [64.4,65.1] (3) \\ 
\emph{restcountries} & 65.5 [63.7,68.5] (5) & {\bf 76.1 [76.1,76.1] (1)} & 3.5 [3.5,3.5] (7) & 50.6 [50.6,50.6] (6) & 73.0 [71.5,74.2] (4) & 75.4 [73.5,76.6] (2) & 73.9 [72.4,75.0] (3) \\ 
\emph{scout-api} & 18.1 [18.1,18.1] (5) & {\bf 36.7 [32.8,41.1] (1)} & 12.0 [12.0,12.0] (6.5) & 26.5 [26.4,26.6] (2) & 12.0 [12.0,12.0] (6.5) & 23.0 [21.4,23.8] (4) & 23.0 [22.1,25.1] (3) \\ 
\midrule 
Average  & 41.9 (4.6) & 56.8 (1.6) & 35.4 (5.8) & 34.6 (5.1) & 33.6 (5.0) & 45.4 (3.9) & 54.5 (1.9) \\ 
\bottomrule 
\end{tabular} 

%% file: generated_files/tableUTests_1h.tex
\begin{tabular}{ l  r r r r r r    }\\ 
\toprule 
SUT  &  \bBOXRT &  \RestCT &  \Restler &  \RestTest &  \RestTestGenVT &  \Schemathesis \\ 
\midrule 
\emph{cyclotron} & {\bf 0.001} & {\bf 0.001} & {\bf 0.001} & {\bf 0.001} & {\bf 0.001} & {\bf 0.001} \\ 
\emph{disease-sh-api} & {\bf 0.001} & {\bf 0.001} & {\bf 0.001} & {\bf 0.001} & {\bf 0.001} & {\bf 0.001} \\ 
\emph{js-rest-ncs} & {\bf 0.001} & 0.721 & {\bf 0.001} & {\bf 0.001} & {\bf 0.003} & {\bf 0.001} \\ 
\emph{js-rest-scs} & {\bf 0.001} & {\bf 0.001} & {\bf 0.001} & {\bf 0.001} & {\bf 0.001} & {\bf 0.001} \\ 
\emph{realworld-app} & {\bf 0.001} & {\bf 0.001} & {\bf 0.001} & {\bf 0.001} & {\bf 0.001} & {\bf 0.001} \\ 
\emph{spacex-api} & {\bf 0.001} & {\bf 0.001} & {\bf 0.001} & {\bf 0.001} & {\bf 0.001} & {\bf 0.001} \\ 
\hline 
\emph{catwatch} & {\bf 0.001} & {\bf 0.001} & {\bf 0.001} & {\bf 0.001} & {\bf 0.001} & 0.909 \\ 
\emph{cwa-verification} & {\bf 0.001} & {\bf 0.001} & {\bf 0.001} & {\bf 0.001} & {\bf 0.001} & 0.434 \\ 
\emph{features-service} & {\bf 0.001} & {\bf 0.001} & {\bf 0.001} & {\bf 0.001} & {\bf 0.001} & {\bf 0.001} \\ 
\emph{gestaohospital-rest} & {\bf 0.001} & {\bf 0.001} & {\bf 0.001} & {\bf 0.001} & {\bf 0.001} & {\bf 0.001} \\ 
\emph{ind0} & NaN & {\bf 0.001} & {\bf 0.001} & NaN & NaN & NaN \\ 
\emph{languagetool} & {\bf 0.001} & {\bf 0.001} & {\bf 0.001} & {\bf 0.001} & {\bf 0.001} & {\bf 0.001} \\ 
\emph{ocvn-rest} & {\bf 0.001} & {\bf 0.001} & {\bf 0.001} & {\bf 0.001} & {\bf 0.001} & {\bf 0.033} \\ 
\emph{proxyprint} & {\bf 0.001} & {\bf 0.001} & {\bf 0.001} & {\bf 0.001} & {\bf 0.001} & {\bf 0.001} \\ 
\emph{rest-ncs} & {\bf 0.001} & {\bf 0.001} & {\bf 0.001} & {\bf 0.001} & 0.179 & {\bf 0.001} \\ 
\emph{rest-news} & {\bf 0.001} & {\bf 0.001} & {\bf 0.001} & {\bf 0.001} & {\bf 0.001} & 0.069 \\ 
\emph{rest-scs} & {\bf 0.001} & {\bf 0.001} & {\bf 0.001} & {\bf 0.001} & {\bf 0.039} & {\bf 0.007} \\ 
\emph{restcountries} & {\bf 0.001} & {\bf 0.001} & {\bf 0.001} & {\bf 0.001} & 0.417 & {\bf 0.001} \\ 
\emph{scout-api} & {\bf 0.001} & {\bf 0.001} & {\bf 0.001} & {\bf 0.001} & {\bf 0.001} & {\bf 0.001} \\ 
\bottomrule 
\end{tabular} 

%% file: generated_files/tableComparisonWB_1h.tex
\begin{tabular}{ l rrrr  rrrr}\\ 
\toprule 
SUT & \multicolumn{4}{c}{Line Coverage \%} & \multicolumn{4}{c}{\# Detected Faults} \\ 
    & RS & WB  & $\hat{A}_{12}$ & p-value  & RS & WB  & $\hat{A}_{12}$ & p-value \\ 
\midrule 
\emph{cyclotron} & 31.2 & 31.3 & 0.69 & 0.158 & 32.2 & 30.9 & {\bf 0.13} & 0.004 \\ 
\emph{disease-sh-api} & 18.1 & 19.5 & {\bf 1.00} & $0.001$ & 30.0 & 36.2 & {\bf 1.00} & $0.001$ \\ 
\emph{js-rest-ncs} & 60.7 & 83.2 & {\bf 1.00} & $0.001$ & 6.0 & 6.0 & 0.50 & 1.000 \\ 
\emph{js-rest-scs} & 60.9 & 76.2 & {\bf 1.00} & $0.001$ & 1.0 & 1.0 & 0.50 & 1.000 \\ 
\emph{realworld-app} & 24.6 & 24.4 & {\bf 0.03} & $0.001$ & 25.0 & 30.5 & {\bf 1.00} & $0.001$ \\ 
\emph{spacex-api} & 41.5 & 41.5 & 0.58 & 0.540 & 52.2 & 50.9 & 0.32 & 0.170 \\ 
\hline 
\emph{catwatch} & 41.1 & 50.0 & {\bf 1.00} & $0.001$ & 18.0 & 23.7 & {\bf 1.00} & $0.001$ \\ 
\emph{cwa-verification} & 40.4 & 46.9 & {\bf 1.00} & $0.001$ & 4.0 & 4.0 & 0.50 & 1.000 \\ 
\emph{features-service} & 68.8 & 81.8 & {\bf 1.00} & $0.001$ & 22.2 & 31.4 & {\bf 1.00} & $0.001$ \\ 
\emph{gestaohospital-rest} & 39.4 & 39.5 & {\bf 0.72} & 0.024 & 15.0 & 22.0 & {\bf 1.00} & $0.001$ \\ 
\emph{ind0} & 8.4 & 18.8 & {\bf 1.00} & $0.001$ & 1.0 & 48.1 & {\bf 1.00} & $0.001$ \\ 
\emph{languagetool} & 28.9 & 39.5 & {\bf 0.93} & $0.001$ & 16.6 & 12.8 & 0.55 & 0.731 \\ 
\emph{ocvn-rest} & 35.3 & 24.8 & {\bf 0.00} & $0.001$ & 279.9 & 257.9 & {\bf 0.00} & $0.001$ \\ 
\emph{proxyprint} & 53.3 & 51.6 & 0.28 & 0.120 & 84.1 & 86.3 & 0.59 & 0.531 \\ 
\emph{rest-ncs} & 61.9 & 93.0 & {\bf 1.00} & $0.001$ & 5.0 & 6.0 & {\bf 1.00} & $0.001$ \\ 
\emph{rest-news} & 55.4 & 66.5 & {\bf 1.00} & $0.001$ & 5.0 & 7.8 & {\bf 1.00} & $0.001$ \\ 
\emph{rest-scs} & 63.3 & 86.2 & {\bf 1.00} & $0.001$ & 1.0 & 12.0 & {\bf 1.00} & $0.001$ \\ 
\emph{restcountries} & 74.9 & 77.1 & {\bf 1.00} & $0.001$ & 2.0 & 2.0 & 0.50 & 1.000 \\ 
\emph{scout-api} & 54.7 & 53.4 & 0.37 & 0.346 & 94.0 & 88.6 & {\bf 0.14} & 0.008 \\ 
\midrule 
Average  & 45.4 & 52.9 & 0.77 &  & 36.5 & 39.9 & 0.67 &  \\ 
\bottomrule 
\end{tabular}